\documentclass[twocolumn,showpacs,preprintnumbers,prb]{revtex4-1}

\usepackage{graphicx}
\usepackage{dcolumn}
\usepackage{bm}
\usepackage{amsmath}
\usepackage{amssymb}
\usepackage{hyperref}

\newcommand{\br}{\mathbf{r}}

\newcommand{\be}{\mathbf{e}}

\newcommand{\bv}{\mathbf{v}}

\newcommand{\bM}{\mathbf{M}}
\newcommand{\bR}{{\mathbf{R}}}
\newcommand{\bB}{\mathbf{B}}
\newcommand{\bD}{\mathbf{R}_d}
\newcommand{\bG}{\mathbf{G}}
\newcommand{\cD}{\mathcal{D}}
\newcommand{\cG}{\mathcal{G}}

\begin{document}
\title{Capturing of a Magnetic Skyrmion with a Hole}
\author{Jan M\"uller$^1$}
\email{jmueller@thp.uni-koeln.de}
\author{Achim Rosch$^{1}$}
\affiliation{$^1$ Institut f\"ur Theoretische Physik, Universit\"at zu K\"oln, D-50937 Cologne, Germany}
\date{\today}

\begin{abstract}
Magnetic whirls in chiral magnets, so-called skyrmions, can be manipulated by ultrasmall current densities.
Here we study both analytically and numerically the interactions of a single skyrmion in two dimensions with a
small hole in the magnetic layer.
Results from micromagnetic simulations are in good agreement with effective equations of motion obtained from a generalization of the Thiele approach. Skyrmion-defect interactions are described by an effective  potential with both repulsive and attractive components.
For small current densities a previously pinned skyrmion stays pinned whereas an unpinned skyrmion moves around the impurities and never gets captured.
For higher current densities, $j_{c1} < j < j_{c2}$, however, single holes are able to capture moving skyrmions. The maximal cross section is proportional to the skyrmion radius and to $\sqrt{\alpha}$, where $\alpha$ is the Gilbert damping.
For $j > j_{c2}$ all skyrmions are depinned. Small changes of the magnetic field strongly change the pinning properties, one can even reach a regime without pinning, $j_{c2}=0$.
We also show that a small density of holes can effectively accelerate the motion of the skyrmion and introduce a Hall effect for the skyrmion. 
\end{abstract}

\maketitle

\section{Introduction}

\begin{figure}[htbp]
  \centering
  \begin{minipage}[b]{0.47\textwidth}
    \centering
    \includegraphics[width=\textwidth]{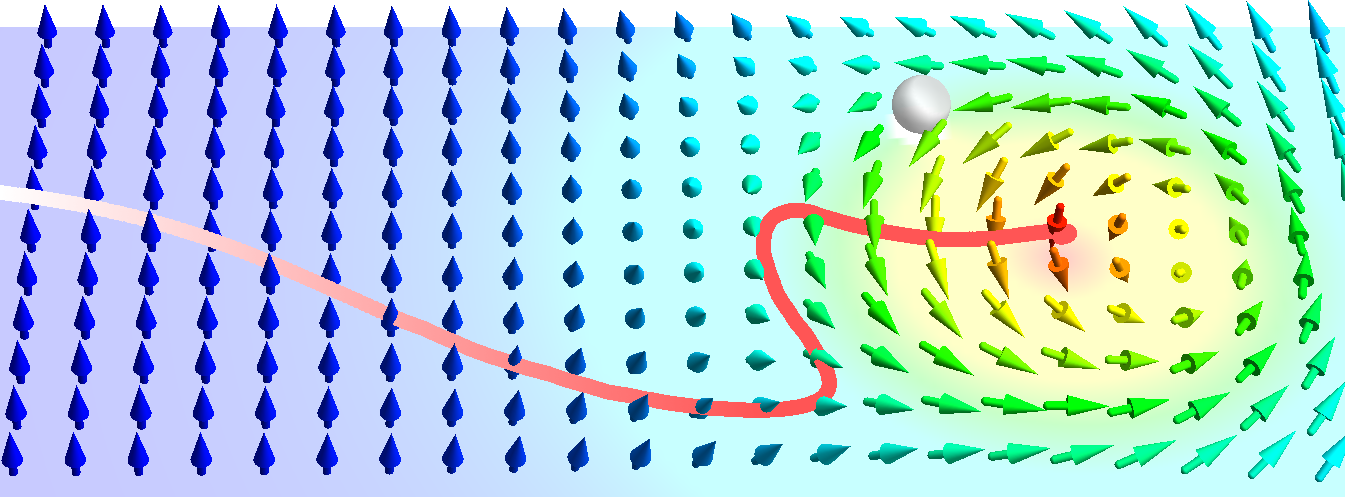}
    \caption{Snapshot of a micromagnetic simulation of skyrmion driven by a current ($D=0.3 J/a$, $\mu B=0.09 J/a^2$, $\bv_s =0.001 aJ\mathbf{e}_x$, and $\alpha=\beta=0.4$) in the presence of a single vacancy: a missing spin (grey sphere). 
    The trajectory of the skyrmion center is indicated by a red line.}
    \label{fig1}
  \end{minipage}
\end{figure}

Topologically stable magnetic whirls, so-called skyrmions, have recently gained a lot of attention
both due to their interesting physical properties and their potential for applications. 
A single skyrmion is shown in Fig.~\ref{fig1}.  
A skyrmion is a smooth magnetic configuration where the spin direction winds once around the unitsphere. 
This implies that the spin configuration is topological protected and can unwind only by creating singular spin configurations \cite{lit:monopoles, lit:26}.
In bulk chiral magnets, lattices of skyrmions are stabilized by Dzyaloshinskii-Moriya interactions and thermal fluctuations in a small temperature and field regime \cite{lit:muehlbauer}.
In films of chiral magnets they occur as a stable phase \cite{lit:8} in a wide range of temperatures in the presence of a stabilizing field. 
Single skyrmions are metastable in an even broader regime of parameters \cite{lit:8}. 
They have been observed in a wide range of materials, including insulators~\cite{lit:10}, doped semiconductors~\cite{lit:7,lit:8} and metals~\cite{lit:muehlbauer,lit:5,lit:6}, with sizes ranging from a few nanometers up to micrometers and from cryogenic temperatures almost up to room temperature \cite{lit:6}.
In bilayer PdFe films on Ir substrates, single nanoscale skyrmions have been written using the current through a magnetic tip \cite{lit:wiesendanger}.

Due to their efficient coupling to electrons by Berry phases and the smoothness of the magnetic texture, skyrmions can be manipulated by extremely small electric current densities \cite{lit:18,ultrasmall,lit:universal,lit:23,lit:24}. 
Therefore the potential exists to realize new types of memory or logic devices based on skyrmions \cite{lit:fert-nanostructures,lit:26}. 
Several studies have therefore investigated the dynamics of skyrmions in nanostructures and their creation at defects \cite{lit:constricted,lit:fert-nanostructures}.

In this paper, we investigate how a single defect affects the dynamics of a single skyrmion in a magnetic film, see e.g. Fig.~\ref{fig1}. 
As an example of a defect we consider a vacancy, i.e., a single missing spin, or more general a hole in the magnetic film with radius small compared to the skyrmion radius.  
This problem is of interest for at least two reasons. 
First, this is perhaps the most simple example of a nanostructure which can interact with the skyrmion. 
As we will show, one can use such defects to control the capturing and release of skyrmions via the magnetic field and the applied current density. 
Second, defects are always present in real materials. 
As long as the typical distance of defects is small compared to the skyrmion radius, the effects of a finite density of defects can be computed from the solution of the single-defect problem. 
The influence of a finite defect density on a lattice of skyrmions has been studied using micromagnetic simulations by Iwasaki, Mochizuki and Nagaosa, Ref.~\citenum{lit:universal}. 
Interestingly, they observed in their simulations that skyrmions move efficiently around defects. 
While a different type of defect (enhanced easy axis anisotropy) was considered by them, a similar phenomenon will also be of importance for our study.

Besides the use of micromagnetic simulations, the main theoretical tool will be the analysis of effective equations of motion for the center of the skyrmion. 
Thiele \cite{lit:Thiele} pioneered the approach to project the effective equations of motion on the translational mode of a magnetic texture. 
This approach has also been successfully applied to skyrmions and skyrmion lattices \cite{lit:19,lit:20,lit:universal,lit:constricted}. 
Here, we combine this approach with microscopic evaluations of an effective potential describing the defect-skyrmion interaction. 
The resulting effective equation of motion for the skyrmion accurately reproduces the results of the micromagnetic simulation and allows for an analytical analysis of the skyrmion dynamics.

In the following, we will first introduce the model, derive the effective potential and resulting equations of motion,
and use them to investigate how skyrmions are captured, released and deflected by a single defect. Finally, we study the effects of a finite, but low density of defects.

\section{The model}
To describe the magnetization of the system we use classical Heisenberg spins $\mathbf{M}(\mathbf{r})$ with uniform length $\|\mathbf{M}(\mathbf{r})\| \!=\! 1$ on a square lattice.
The corresponding free energy functional in the continuum reads
\begin{equation}
	F[\mathbf{M}] \!=\! \!\!
	   \int\! \!\text{d}^2 r \,
	     \frac{J}{2} \left[\nabla \mathbf{M} (\mathbf{r}) \right]^2 
	    \!+\! D \mathbf{M} (\mathbf{r}) \!\cdot\! \left[ \nabla\! \!\times\! \mathbf{M} (\mathbf{r}) \right] 
	     -\mu  \mathbf{B} \!\cdot\! \mathbf{M} (\mathbf{r})  \text{,}
\label{eq:FreeEnergy}
\end{equation}
including the ferromagnetic coupling $J$, Dzyaloshinskii-Moriya interaction $D$ and the Zeeman interaction with the magnetic field $\mathbf{B} = (0, 0, B)$. 
$\mu$ is the magnetization per area. 
For a single spin $1/2$ per square unit cell with lattice constant $a_0$ and $g$ factor $g=2$ one has, for example,  $\mu=\mu_B/a_0^2$.

On a square lattice we use the following discretized version 
%
%
\begin{eqnarray}
F[\bM] &=& -J \sum_{\br} \bM_\br \cdot \left( \bM_{\br+a\be_x} + \bM_{\br+a\be_y} \right)\nonumber\\
&-& D a \!\sum_{\br} \left( \bM_\br \times \bM_{\br+a\be_x} \!\!\cdot \be_x + \bM_\br \times \bM_{\br+a\be_y} \!\!\cdot \be_y \right)\nonumber\\
&-& \bB \mu a^2 \cdot \sum_\br \bM_\br \text{,}
\label{eq:FreeEnergy2}
\end{eqnarray}
%
%
where $\be_x$ and $\be_y$ are unit vectors in the $x$ and $y$ direction, respectively. The lattice constant $a$ and the interaction strength $J$ are set to $1$ in the following. 
If not otherwise stated, we use $D=0.3 J/a$ and $\mu B=0.09 J/a^2$. 
For these parameters the ground state is ferromagnetic.
Hence the single skyrmion is a topologically protected, metastable excitation.
A vacancy at position $\bD$ is created by setting the magnetization $\mathbf{M}$ at this site to zero.

The micromagnetic dynamics of each spin in the presence of an electric current density $j$ are described by the Landau-Lifshitz-Gilbert (LLG) equation \cite{lit:LLG1,lit:LLG2,lit:LLG3}.
In the continuum case the LLG equation reads
\begin{eqnarray}
\left[\partial_\text{t} \!+\! \left( \mathbf{v}_s \!\cdot\! \nabla \right)\right]\! \mathbf{M}= &-&\gamma \mathbf{M} \times \mathbf{B_\text{eff}} \nonumber\\
&+& \alpha \mathbf{ M} \!\times\! \left[ \partial_\text{t} \mathbf{M} + \frac{\beta}{\alpha} \!\left( \mathbf{v}_s \!\cdot\! \nabla \right) \mathbf{M} \right] \text{,}
\label{eq:LLG}
\end{eqnarray}
where $v_s$ is the drift velocity of spin currents which is directly proportional to the current density $j$ and $\gamma=g \mu_B/\hbar$ is the gyromagnetic ratio. 
Note that we set $v_s, \alpha$ and $\beta$ to a constant value, not taking into account that depending on the microscopic realization of the defect, they might be modified in proximity of the defect. 
At least for defects small compared to the skyrmion radius and sufficiently small currents, this approximation is justified as the forces on the skyrmions add up from all parts of the skyrmion (see below). 
The (very weak) effects of changes to the  current pattern around a notch in a nanowire have been studied in Ref.~\citenum{lit:constricted}.
The effective magnetic field is given by  $\mathbf{B_\text{eff}} = - \frac{\delta F[\mathbf{M}]}{\mu \delta \mathbf{M}}$. $\alpha$ and $\beta$ are phenomenological damping terms. 
Note that $\alpha=\beta$ is a special point as in this case the magnetic texture drifts with the current as long as no defects are present, $\bM(\br,t)=\bM(\br-\mathbf{v}_s t)$. 
In our lattice model we rewrite Eq.~(\ref{eq:LLG}) using  $\partial_i \mathbf{M}(\mathbf{r}) = \frac{1}{2 a} \left( \mathbf{M}_{\br +a  \be_{i}} - \mathbf{M}_{\br -a  \be_{i}} \right)$.

\section{Effective dynamics of skyrmions}

\subsection{Generalized Thiele approach}

The LLG equation describes the movement of every magnetic moment in the system.
As we do not want to describe every spin but the movement of the skyrmion center, which is a collective movement of spins, we apply the so-called Thiele approach \cite{lit:Thiele}. 
Originally, this approach is based on the approximation that the skyrmion is a completely rigid object. 
While this approximation fails in the presence of a local defect, we will show that one can nevertheless use this approach if one performs a microscopic calculation of the potential $V(\br)$ describing the forces between skyrmion and defect.

Our goal is to derive an equation of motion for the center $\bR$ of the skyrmion ($\bR$ is defined below), which takes into account deformations of the skyrmion. 
If the motion of the skyrmion is sufficiently slow, we expect that for each fixed $\bR$ the skyrmion configuration is in a local minimum of the energy. 
We therefore approximate
\begin{eqnarray}
\bM(\br,t) \approx \bM_0(\,\bR(t)-\bD\,,\,\br-\bR(t)\,)  \text{.}
\label{m0}
\end{eqnarray}
The magnetic configuration $\bM_0$ depends on the distance of skyrmion center $\bR(t)$ and defect position $\bD$ and is determined from the condition that 
\begin{align}
V(\bR-\bD) &= F[\bM_0(\,\bR-\bD\,,\,\br-\bR\,)]-F_0 \nonumber \\
&=\min_{\bR-\bD \ \text{fixed}} F[\bM(\br)]-F_0
\label{V}
\end{align}
is at a local minimum for {\em fixed} distance of skyrmion and defect, $\bR-\bD$. 
$V(\bR-\bD)$ is the effective potential describing the skyrmion-defect interaction. The offset $F_0$ is chosen such that $V(R\to \infty)=0$. 
Note that the standard Thiele approach {\em neglects} the deformation of the skyrmion, i.e., the dependence of $\bM$ on $\bR-\bD$.

\begin{figure}[t]
  \centering
  \begin{minipage}[b]{0.47\textwidth}
    \centering
    \includegraphics[width=\textwidth]{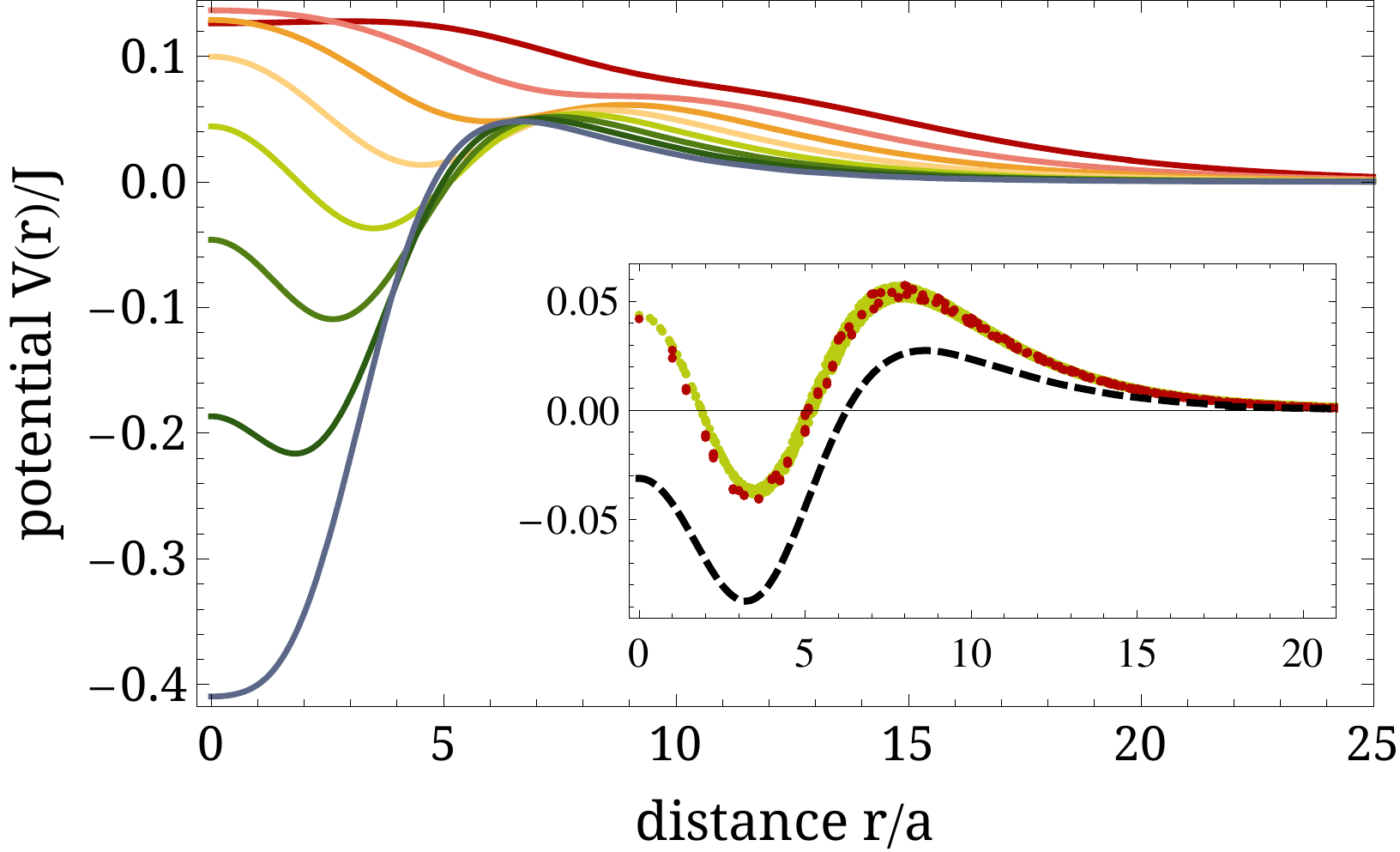}
    \caption{Potential $V(\bR-\bD)$ of the skyrmion-hole interaction as a function of distance shown for $D=0.3 J/a$ and various magnetic fields from $\mu B=0.05 J/a^2$ (red) to $\mu B=0.12 J/a^2$ (blue) in steps of $\Delta \mu B =0.01 J/a^2$.
    Inset: Raw data used to calculate the smoothened potential shown in the main figure. 
    The dark red (light green) data has been obtained for $\mu B =0.09J/a^2$ using the first and second algorithm described in the text. 
    The spread in each curve arises as on the square lattice the potential does not only depend on the distance from the defect but also has a tiny angular dependence. 
    For comparison, we also show the estimate for the potential which is obtained when deformations of the skyrmion are ignored (dashed line).}
    \label{fig2}
  \end{minipage}
\end{figure}
To calculate $\bM_0$ and $F[\bM_0]$ numerically, we have used two different methods.
In the case that $\bR$ is located on one of the lattice sites, we fix the position $\bR$ of the skyrmion by setting the magnetization at $\br=\bR$ to $(0,0,-1)$, opposite to the ferromagnetic background. 
This approach is similar to the method used in Ref.~\citenum{lit:24} to numerically calculate the potential of the skyrmion-skyrmion interaction.
In a second approach, we first compute the skyrmion configuration $\bM_{c}(\br-\bR)$ in the clean system without a defect. 
To determine the energy minimum in the presence of the defect for fixed $\bR$, we minimize $ F[\bM(\br)]$ with the boundary condition that $\bM(\br)=\bM_{c}(\br-\bR)$ for $|\br-\bD|>r_0$. 
It turns out that this procedure rapidly converges with $r_0$ and $r_0=4.5 a$ gives accurate results in the considered parameter range.
The results for $V(\bR-\bD)$ determined from the two methods are almost identical, see inset of Fig.~\ref{fig2}.

In Fig.~\ref{fig2} the resulting potentials $V(|\bR-\bD|)$ are shown.
In the continuum model, Eq.~(\ref{eq:FreeEnergy}), the effective potential depends only on the distance of skyrmion and defect, $|\bR-\bD|$, whereas in the lattice there is a small angular dependence (raw data is shown in the inset of Fig.~\ref{fig2}).
For simplicity, we average over this angular dependence.
We fit an exponential law for very large $|\bR-\bD|$ and interpolate the curve by a polynomial otherwise. 
The shape of the potential not only quantitatively but also qualitatively depends on the strenght of the magnetic field, which will be important for the following discussion.

To derive an effective equation of motion for $\bR(t)$, we proceed as follows \cite{lit:Thiele}. 
First, both sides of the  LLG equation are multiplied by $\frac{\mu}{\gamma} \bM \times$ such that  $\mu \mathbf{B_\text{eff}} = - \frac{\delta F[\mathbf{M}]}{\delta \mathbf{M}}$ is isolated (using that $\mathbf{B_\text{eff}}$ can be chosen to be perpendicular to $\bM$). 
Second, $\bM$ is replaced by $\bM_0$ defined in Eq.~(\ref{m0}). 
Third, to project onto the translational mode in direction $i$ the resulting equation is multiplied by $\frac{d \bM_0}{d R_i}$ and integrated over space.

The resulting differential equation for $\bR(t)$, the generalized Thiele equation, can be written in the following form
\begin{align}
-\frac{d V}{d \bR} =\,\,&
 \bG_\bR \times \left( \dot{\bR} -  \bv_s \right) + \delta \cG_\bR \cdot \bv_s \nonumber \\
& + \cD_\bR \cdot \left( \alpha \dot{\bR} - \beta \bv_s \right) + \beta \delta \cD_\bR \cdot \bv_s \text{,}
\label{thiele}
\end{align}
where the potential $V$, the gyrocoupling $\bG_\bR$ and the matrices  $\delta \cG_\bR$, $\cD_\bR$ and $\delta \cD_\bR$ are functions of the distance from the defect, $\bR-\bR_d$. 
$V$ is defined in Eq.~(\ref{V}), the other terms are determined from 
\begin{align}
 \left( \bG_\bR \right)_i &= s\,  \epsilon_\text{ijk}  \!\int\!\! \mathrm{d}^2 r \,\, \frac{1}{2} \bM_0 \cdot \left( \frac{d \bM_0}{d R_j} \times  \frac{d \bM_0}{d R_k} \right) \label{eq:defG} \\
 \left( \cD_\bR \right)_{ij} &= s \!\int\!\! \mathrm{d}^2 r\frac{d \bM_0}{d R_i} \cdot  \frac{d \bM_0}{d R_j} \label{eq:defD} \\
 \left( \delta \cG_\bR \right)_{ij} &= s \!\int\!\! \mathrm{d}^2 r\,\, \mathbf{M_0} \cdot \left( \frac{d \bM_0}{d R_j } \times\! \left( \frac{d \bM_0}{d R_i}+\frac{d \bM_0}{d r_{i}} \right)\!\!\right)\\
 \left( \delta \cD_\bR \right)_{ij} &= s \!\int\!\! \mathrm{d}^2 r\,\, \frac{d \bM_0}{d R_{i}} \cdot   \left( \frac{d \bM_0}{d R_j}+\frac{d \bM_0}{d r_{j}} \right) \text{,}
\end{align}
where we included the spin density
\begin{equation}
s= \frac{\mu}{\gamma} \text{,}
\end{equation}
e.g., $s=\hbar/(2 a_0^2)$ for a single spin 1/2 in a unit cell of length $a_0$.
Note that some of the derivatives are with respect to the skyrmion position $\bR$ and further that the combination  $\frac{d \bM_0}{d R_i}+\frac{d \bM_0}{d r_{i}}=-\frac{d \bM_0}{d R_{d,i}}$ describes the change of the skyrmion configuration when only the position of the defect changes.

If the deformation of the skyrmion (and therefore the derivatives $\frac{d}{d R_{d,i}}$) are ignored, then the correction terms $\delta \cG_\bR$ and $\delta \cD_\bR$ vanish and one can replace $\frac{d}{d R_i}$ by $-\frac{d}{d r_i}$ to recover the Thiele equation in the standard form.
Within this approximation, the gyrocoupling $\bG_\bR=\bG$ is in the continuum limit topologically quantized to a multiple of $4 \pi M$ ($M$ is the magnetization per unit cell set to 1 within our conventions). 
This is, however, {\bf not} the case if the dependence of $\bM_0$ on $\bR-\bR_d$ is taken into account.

The most important effect of the deformation is that they strongly modify the effective potential $V(\bR-\bD)$, as is shown in the inset of Fig.~\ref{fig2}. 
Taking into account the adjustment of the magnetic texture to the defect is important as it gives rise to corrections of order 1.

\begin{figure}[htbp]
  \centering
  \begin{minipage}[b]{0.47\textwidth}
    \centering
    \includegraphics[width=\textwidth]{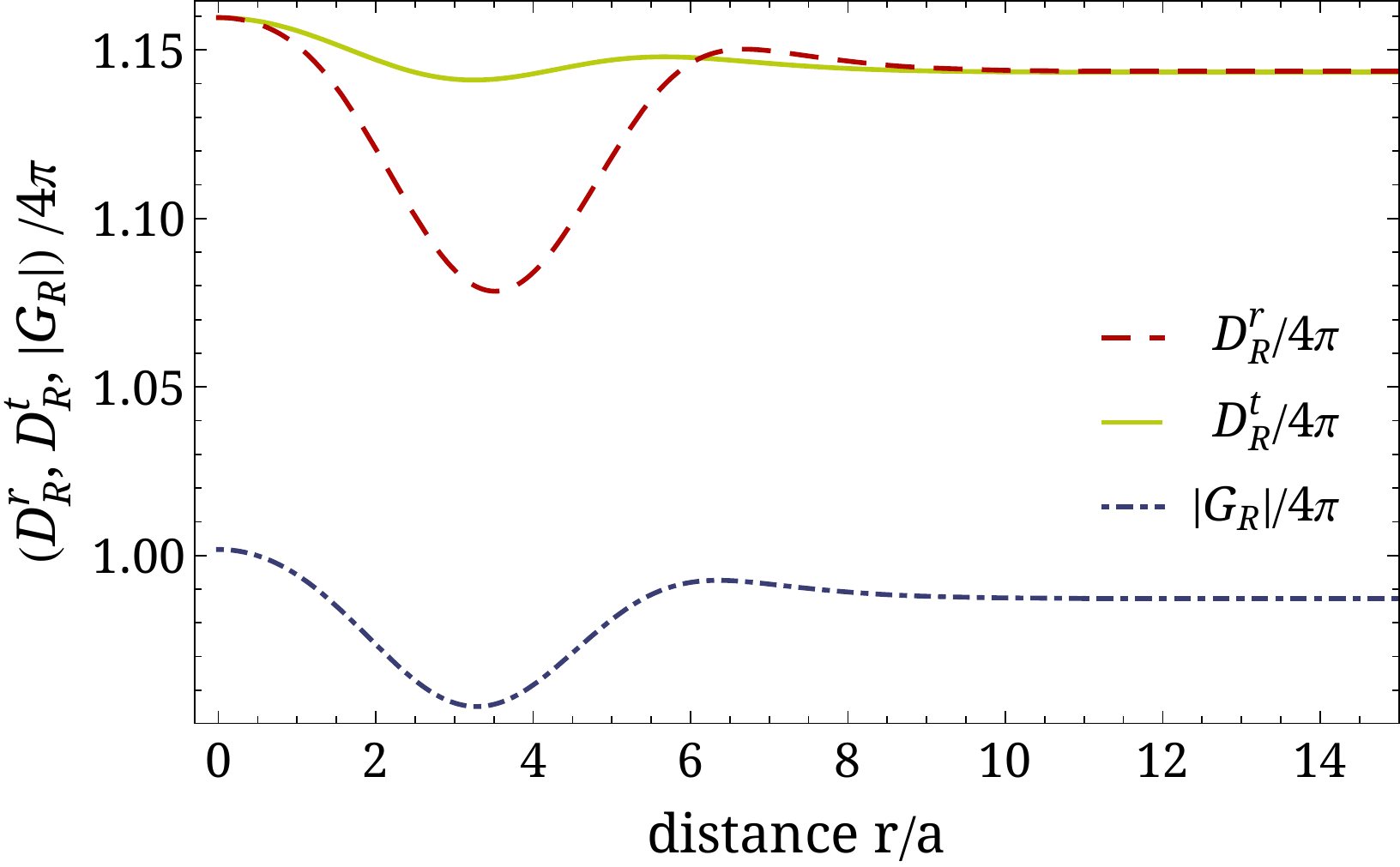}
    \caption{$|\bG_\bR|$, $\cD_\bR ^{\,r}$, and $\cD_\bR ^{\,t}$ shown as a function of the distance from the defect, $r=|\bR-\bD|$, for $D=0.3 J/a$ and magnetic field $\mu B=0.09 J/a^2$.}
    \label{fig3}
  \end{minipage}
\end{figure}

Changes of the gyrocoupling and dissipative tensor are, in general, also of importance when nanostructures lead to a significant deformation of the magnetic texture. 
They are, however, not important for the situation considered in our paper. 
We study the case where the radius $a_d$ of the defect is much smaller than the radius of the skyrmion, $a_s$. 
In this case the deformations affect only a small part of the skyrmion and give therefore only small corrections of order $(a_d/a_s)^{2} \ll 1$ on the right-hand side of the generalized Thiele equation (\ref{thiele}). 
This is shown in Fig.~\ref{fig3}, where $|\bG_\bR|$, $\cD_\bR ^{\,r}$ and $\cD_\bR ^{\,t}$ are shown as a function of the distance from the defect, $|\bR-\bD|$. 
Here $\cD_\bR ^{\,r}=\hat e_r \cdot \cD_\bR \cdot \hat e_r$ and $\cD_\bR ^{\,t}=\hat e_\phi \cdot \cD_\bR \cdot \hat e_\phi$ describe the dissipative tensor projected on the radial and tangential direction, respectively, with $\hat e_r=(\bR-\bD)/|\bR-\bD|$ and $\hat e_\phi=\hat z \times \hat e_r$. 
Far from the defect, one recovers the results predicted by the standard Thiele approach with $\cD_\bR ^{\,r}=\cD_\bR ^{\,t}$ and $|\bG_\bR|=4 \pi$ in the continuum limit, whereas there are small deviations of a few percent when the distance of the defect is of the order of the skyrmion radius. 
Similarly, the corrections arising from $\delta \cD_\bR$ and $\delta \cG_\bR$ are also small.
For the following analysis, we will therefore neglect the modification of the right-hand side of the Thiele equation (\ref{thiele}) using 
\begin{align}
-\frac{d V}{d \bR} =
 \mathbf{G} \times \left( \dot{\bR} -  \bv_s \right)
 + \mathcal{D} \cdot \left( \alpha \dot{\bR} - \beta \bv_s \right)  \text{,}
\label{thieleSimplified}
\end{align}
with space-independent $\bG=\lim_{\bR \to \infty} \bG_\bR$ and $\cD=\lim_{\bR \to \infty} \cD_\bR$ while the modified potential is fully taken into account.

\subsection{Comparison of the generalized Thiele approach and micromagnetic simulations}
 
\begin{figure}[htbp]
  \centering
  \begin{minipage}[b]{0.47\textwidth}
    \centering
    \includegraphics[width=0.49\textwidth]{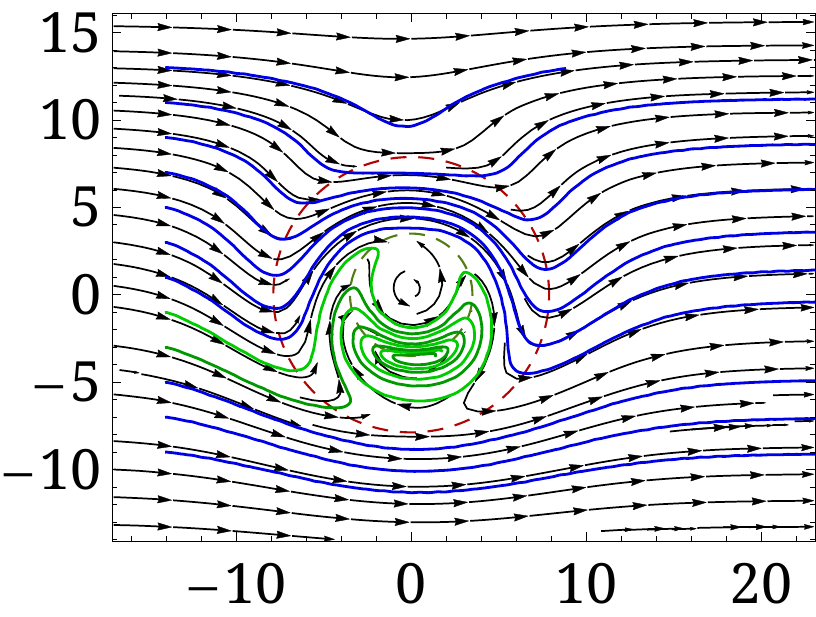}
    \vspace{0.02\textwidth}
    \includegraphics[width=0.49\textwidth]{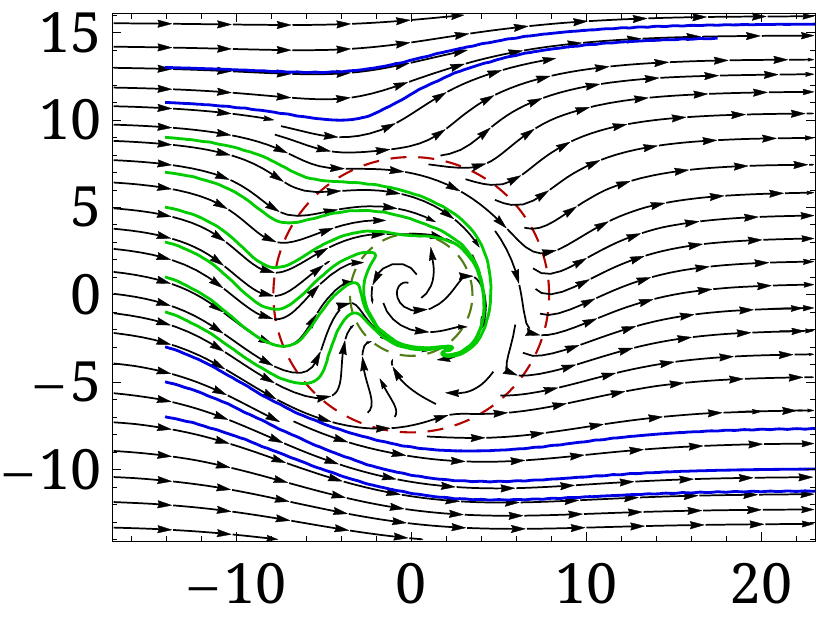}
    \caption{
    Comparison of trajectories of the moving skyrmion obtained from the full micromagnetic simulation (colored) to the results of the effective potential approach (black).
    The vacancy is placed in the origin.
    For better orientation, the green (red) circle indicates the minimum (maximum) of the effective potential.
    Parameters are $D=0.3 J/a$, $\mu B=0.09 J/a^2$, $\bv_s =0.001 aJ\mathbf{e}_x$, and $\alpha=\beta=0.04$ (left), $\alpha=\beta=0.4$ (right).
    }
    \label{fig4}
  \end{minipage}
\end{figure}

Using the numerically determined potential, see Fig.~\ref{fig2}, one can directly calculate the trajectories of the skyrmions using the Thiele equation, Eq.~(\ref{thieleSimplified}). 
In Fig.~\ref{fig4} the trajectories are shown for two values of the damping constants, i.e., $\alpha=\beta=0.4$ and $0.04$. 
The properties of these solutions will be discussed in Sec.~\ref{capturing}.
Here we compare them to full micromagnetic simulations of the system. 
To track the center of the skyrmion $\bR$, we used $\bR \approx \sum_i (M^z_0-M^z_i) \br_i/\sum_i (M^z_0-M^z_i)$ summing only over sites with $M^z_i<M^z_0=-0.3$. 

Comparing the results from the micromagnetic simulations and the simplified Thiele equation, we find that 
the two approaches agree quantitatively with high precision. 
Tiny deviations arise partially from using the simplified Thiele equation (\ref{thieleSimplified}) instead of (\ref{thiele}) neglecting, e.g., the spatial variations of $\mathbf{G}_\bR$ and $\mathcal D_\bR$ shown in Fig.~\ref{fig3}. 
The other source of error is that the motion of the skyrmion leads to internal excitations and emission of spin waves not captured by the generalized Thiele approach. 
These effects are expected to be larger for smaller $\alpha$ but even for $\alpha=0.04$ they give only small quantitative corrections in the considered parameter regime. 
Most importantly, these corrections do not affect the physics of capturing and depinning discussed in the following paragraphs.

\subsection{Scaling of the effective potential and dimensionless units}\label{scalingSec}

Most results presented in this paper have been obtained for a fixed value of the Dzyaloshinskii-Moriya interaction, $D=0.3\,J/a$, and for a defect described by a single missing spin in a two-dimensional lattice. 
As skyrmions are smooth objects, one can use simple scaling relations to determine the effective potential and therefore also the equation of motion and the skyrmion trajectories for other parameters.

Within the continuum theory, the free energy, Eq.~(\ref{eq:FreeEnergy}), in the presence of a defect of radius
$a_d$  is invariant under the transformation $\br \to \br'= \br/\lambda$, $a_d \to a_d'= a_d/\lambda$, $J\to  J' =J$,
$D\to  D' =\lambda  D$, and $\mathbf B  \to \mathbf B' =\lambda^2 \mathbf B$. 
The shape of the skyrmion is thereby determined by the scale invariant dimensionless parameter
\begin{equation}
\zeta=\frac{J \mu B}{D^2} \text{,}
\end{equation}
with  $\zeta'=\zeta$ (called $\kappa^2/Q^2$ in Ref.~\citenum{lit:christoph}).

The scaling invariance implies that the effective potential in the continuum limit can be written as
\begin{equation}
V_{\rm eff}(\br)=J\, V\!\!\left(\zeta,\frac{a_d \mu B}{D},\frac{\br \mu B}{D}\right)
\approx a_d^2 \frac{J \mu^2 B^2}{D^2} V_\zeta\!\left(\frac{\br \mu B}{D}\right) \label{scaling} \text{,}
\end{equation}
where $V$ and $V_\zeta$ are dimensionless scaling functions with dimensionless arguments and $V_\zeta(\xi)=\frac{1}{2}\left. \frac{d^2 V(\zeta,\gamma,\xi)}{d \gamma^2}\right|_{\gamma=0}$. 
For the last expression, we assumed that $a_d$ is much smaller than the skyrmion radius and used that the term linear in $a_d$ vanishes.

\begin{figure}[htbp]
  \centering
  \begin{minipage}[b]{0.47\textwidth}
    \centering
    \includegraphics[width=\textwidth]{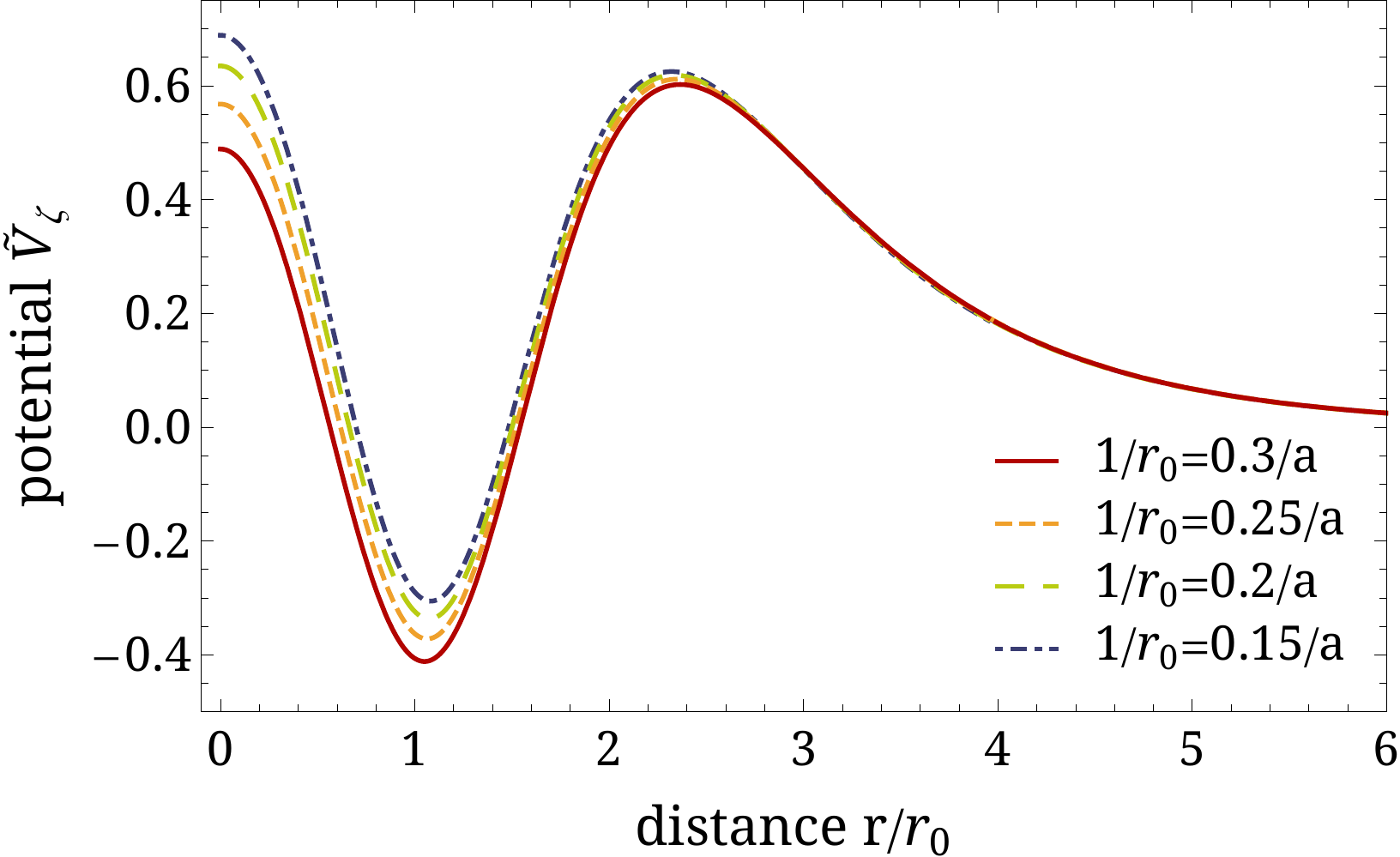}
    \caption{Rescaled potentials, $\tilde V_\zeta = V_{\rm eff}/E_0$ with $E_0= \frac{a^2 J \mu^2 B^2}{D^2}$, see Eq.~(\ref{scaling2}), as a function of the rescaled distance $r/r_0$ with $r_0=D/\mu B$.
    Parameters are $a=a_d=1$ and $\zeta=1$.}
    \label{fig5}
  \end{minipage}
\end{figure}

Strictly speaking, the above derived equation is fully valid in the  limit that the lattice constant $a$ is much smaller than both the skyrmion radius and the defect radius $a_d$. 
But even for a defect given by a single missing spin ($a_d=a$), one can use that the skyrmion radius $a_s$ is much larger than $a$. 
Therefore the dependence of the effective potential on $J, D$ and $\mu B$ can be described as
\begin{equation}
V_{\rm eff}(\br)\approx a^2 \frac{J \mu^2 B^2}{D^2}
 \tilde V_\zeta\!\left(\frac{\br \mu B}{D}\right)  \text{.} \label{scaling2}
\end{equation}
Note that $\tilde V_\zeta(\xi)$ will in general differ from $V_\zeta(\xi)$.
This prediction is confirmed in Fig.~\ref{fig5} which shows $\tilde V_\zeta$ obtained for a single missing spin and various skyrmion sizes obtained by changing $D$ and $B$ such that  $\zeta=1$ remains fixed.

The scaling properties for the skyrmion trajectories can simply be obtained by realizing that both $\bG$ and
$\mathcal D$ are invariant under the scaling transformation as can directly be seen from Eqs.~(\ref{eq:defG}) and (\ref{eq:defD}).
Therefore the (generalized) Thiele equations (\ref{thiele}) or (\ref{thieleSimplified}) are invariant
if one uses for fixed defect radius $a_d$ the scaling $\br \to \br/\lambda$, $D\to \lambda D$, $\bB \to \lambda^2 \bB$ such that $d V_{\rm eff}/d \br \to \lambda^3 d V_{\rm eff}/d \br$
and simultaneously rescales the drift velocity $v_s \to \lambda^3 v_s$ and the time $t \to \lambda^4 t$.

An important consequence of this analysis is that the typical drift velocity $v_s$ of electrons or the corresponding critical current density, $j_c$, to depin a skyrmion from a defect scales with the third power of the inverse skyrmion radius $a_s \sim D/\mu B$
\begin{equation}
j_c \sim v_s \sim {a_s}^{-3} \text{.}
\end{equation}
This is part of the reason why skyrmions can be manipulated by ultrasmall current densities \cite{ultrasmall}.

It is also an interesting question how the potentials change when instead of a single magnetic layer $N_L>1$ layers are considered. 
For a line defect where all spins are removed in a line perpendicular to the surface and for $N_L \gg 1$ one can use that away from the surface the magnetic configuration is translationally invariant in $z$ direction. 
Therefore, the effective potential is  simply given by multiplying $V_{\rm eff}$ by $N_L$.
As also the gyrocoupling and damping matrix scale linearly in $N_L$, the equation of motion for the skyrmion center remains unmodified as long as the phase with a single skyrmion in a ferromagnetic background remains stable. 
Increasing $N_L$ allows to eliminate all effects of thermal fluctuations. 
The situation is more complicated when only a few layers $N_L$ are considered. 
As the properties of the surface and the inner layers are different, $V_{\rm eff}$ cannot simply be computed from the single-layer result.

For the presentation of our results, it is useful to find the minimal set of dimensionless parameters needed to parametrize our results. 
Here it is useful to note, that the dependence on $\beta$ in the effective Thiele equations can be eliminated by parametrizing the effect of the current by the drift velocitiy $\bv_d$ of the skyrmion in the absence of any defect. 
It can be obtained from the equation $\bG \times \bv_s+\beta \cD \bv_s= \bG \times \bv_d+\alpha \cD \bv_d$. 
Further we also define the dimensionless drift velocity $\bv$ by
\begin{eqnarray}
\bv_d&=& \frac{1}{\bG^2+\alpha^2  \cD^2} \left( (\alpha-\beta) \cD \bG \times \bv_s + (G^2+ \alpha \beta \cD^2) \bv_s \right)\nonumber \\
\bv &=& \bv_d  \frac{s D^3}{a^2 J \mu^3 B^3}  \text{.}
\label{eqv}
\end{eqnarray}
In units where all length scales are measured in units of $D/\mu B$ and all times in units of 
$\frac{s D^4}{a^2 J \mu^4 B^4}$ the effective Thiele equation (\ref{thieleSimplified}) takes the form
\begin{align}
-\frac{d \tilde V_\zeta(\bR)}{d \bR} =
 -4 \pi \hat z \times \left( \dot{\bR} - \bv \right)
 +\alpha \cD_\zeta  \left(  \dot{\bR} - \bv \right)  \text{,}
\label{thieleDimensionless}
\end{align}
with $\cD_\zeta = \cD/s$.
Originally, the continuum theory was parametrized by $J$, $D$, $\mu B$, $\alpha$, $\beta$, $\bv_s$, and the size of the defect. 
For a point-like defect, we find that the three dimensionless variables $\zeta$, $\bv$, and $\alpha$ are sufficient to describe all regimes.

\section{Skyrmion Depinning, Capturing and Deflection}\label{capturing}

\subsection{Phase diagram}

When studying the qualitative behavior of the skyrmions when a current is slowly switched on, it is useful to distinguish two initial states, an initially localized skyrmion and a skyrmion approaching the defect from far away.

\begin{figure}[t]
  \centering
  \begin{minipage}[b]{0.47\textwidth}
    \centering
    \includegraphics[width=\textwidth]{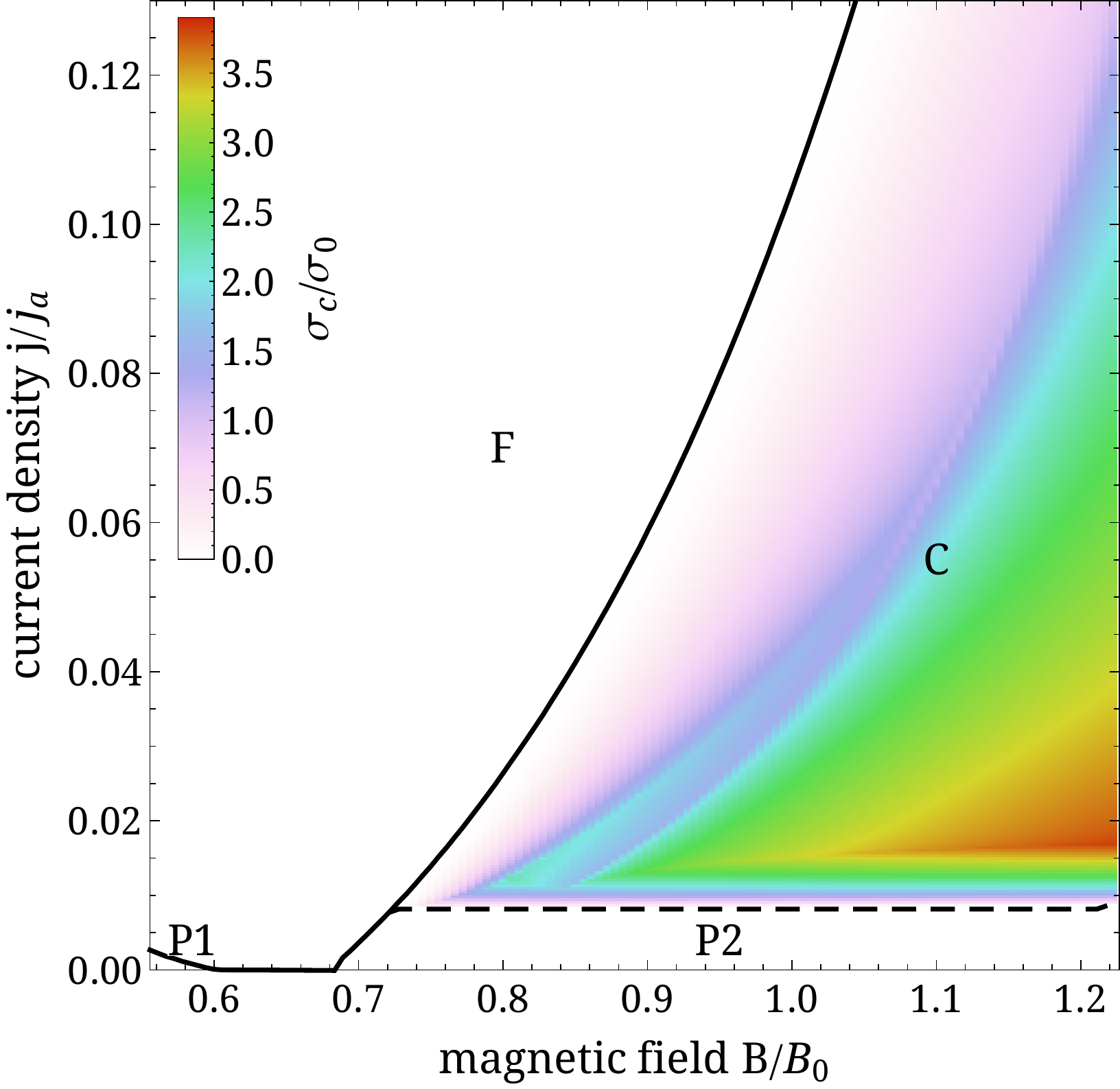}
    \caption{Phase diagram as function of the magnetic field $B/B_0=\zeta$ and current density $j/j_a=v \zeta^3$  for $\alpha=0.1$.
      Here we use the combination $v \zeta^3=v_d \frac{J^2 s}{a^2 D^3}$ as it is independent of the magnetic field.
      The colored area encodes the value for the capturing cross section $\sigma_c/\sigma_0$ with the characteristic length $\sigma_0=D/\mu B$, see Sec.~\ref{sec:CCS}, which is a measure for the efficiency of capturing.
 }   \label{fig6}
  \end{minipage}
\end{figure}

If the skyrmion is initially localized close to the defect and if the potential has a local minimum, it will remain there for small current densities and gets depinned for larger current densities. 
Similarly, a skyrmion approaching the defect from far away can either get captured (green trajectories in Fig.~\ref{fig4}) by the defect or is just deflected (blue trajectories).

An overview over these possibilities is given in the phase diagram, Fig.~\ref{fig6}. 
The solid lines mark the depinning transition. 
Below these lines, in the regimes denoted by P1, P2 and C, an initially localized skyrmion remains localized close to the defect when the current is switched on slowly. 
In P1 the effective potential has a local minimum at $r=0$ while in P2 and C it has a minimum at finite skyrmion-defect distance. 
In the free phase, F, pinning is not possible and all skyrmions move freely. 
At low magnetic fields we find this phase even for zero current density.
Note that we consider only $\zeta>0.56$ as at this point \cite{lit:christoph} the circular symmetric skyrmion becomes unstable towards the formation of a bimeron \cite{lit:merons}.

An unexpected result is that in the pinning regimes P1 and P2 a skyrmion approaching the defect from far apart is {\em not} captured. 
Instead of getting trapped, it moves around the defect and is only deflected. 
This is a consequence of the fact that for long distances the defect-skyrmion potential is repulsive. 
Capturing of approaching skyrmions is only possible in the region C.

\begin{figure}[t]
  \centering
  \begin{minipage}[b]{0.47\textwidth}
   \includegraphics[width=0.48\textwidth]{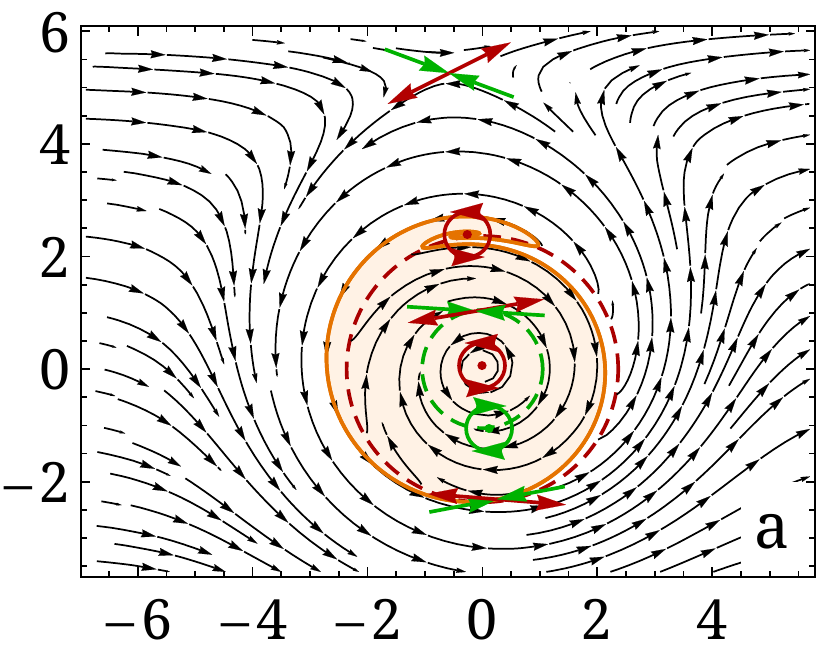} 
    \vspace{0.02\textwidth}
    \includegraphics[width=0.48\textwidth]{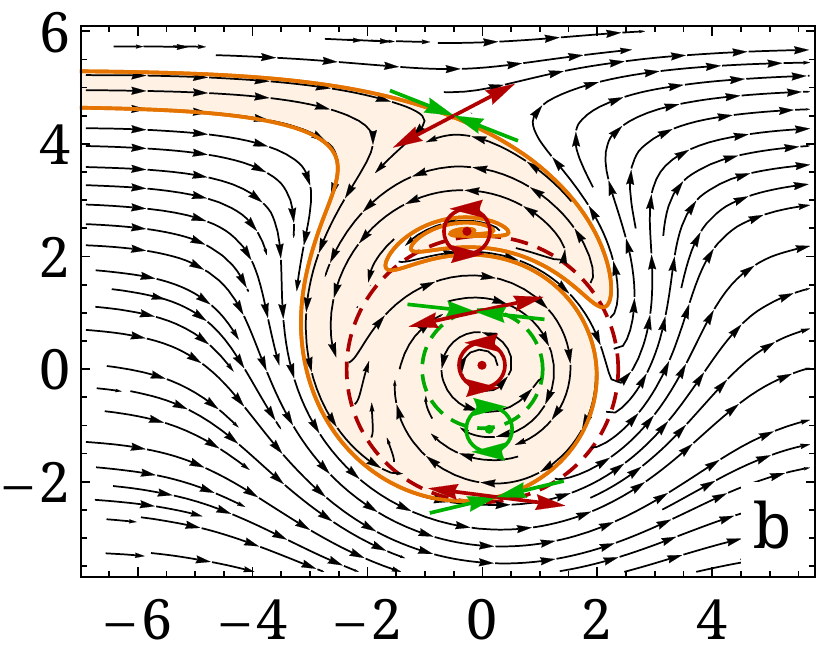} 
    \includegraphics[width=0.48\textwidth]{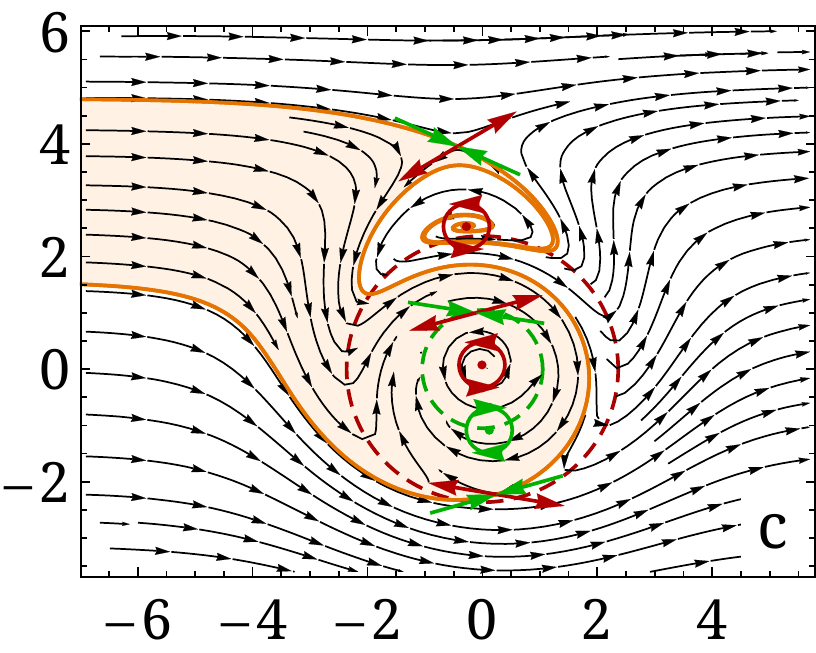} 
    \vspace{0.02\textwidth}
    \includegraphics[width=0.48\textwidth]{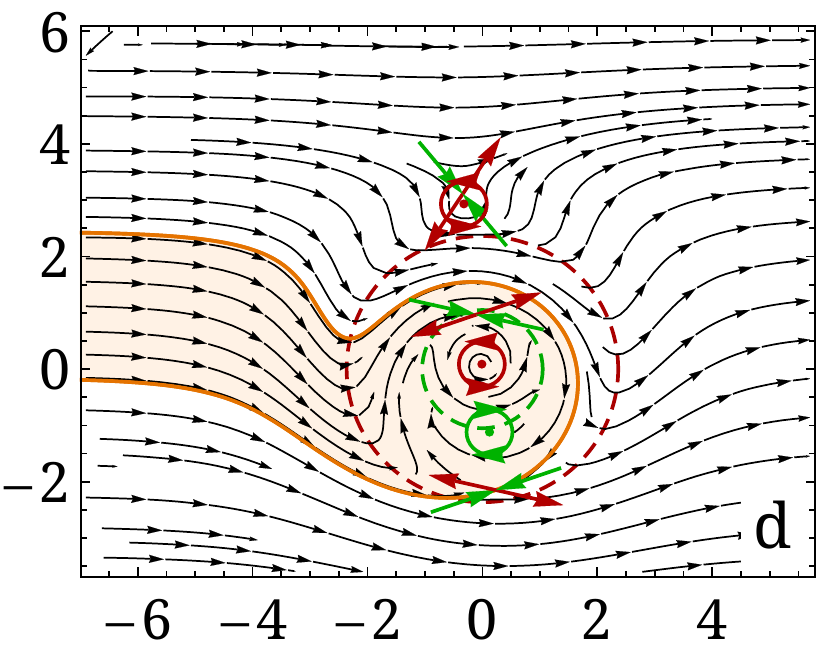} 
    \includegraphics[width=0.48\textwidth]{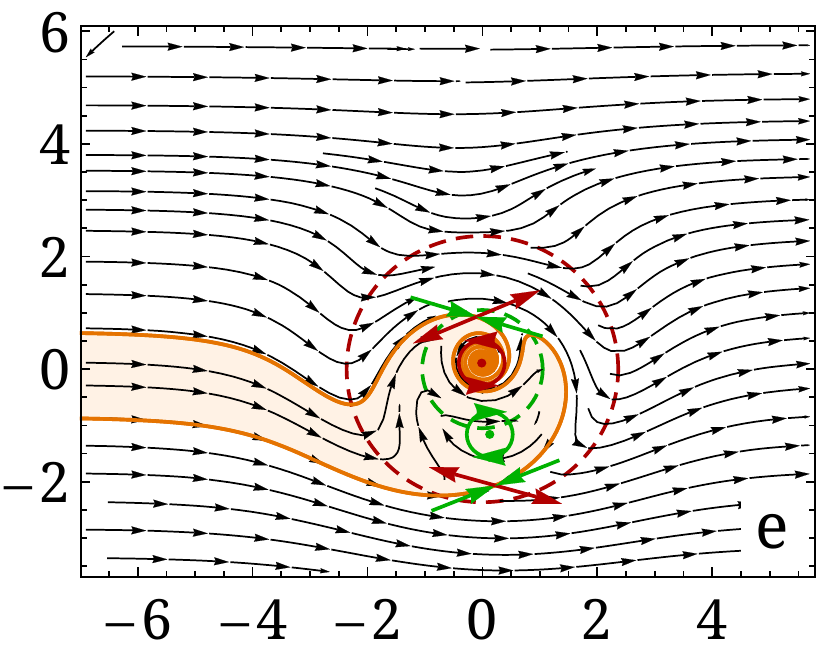} 
    \vspace{0.02\textwidth}
    \includegraphics[width=0.48\textwidth]{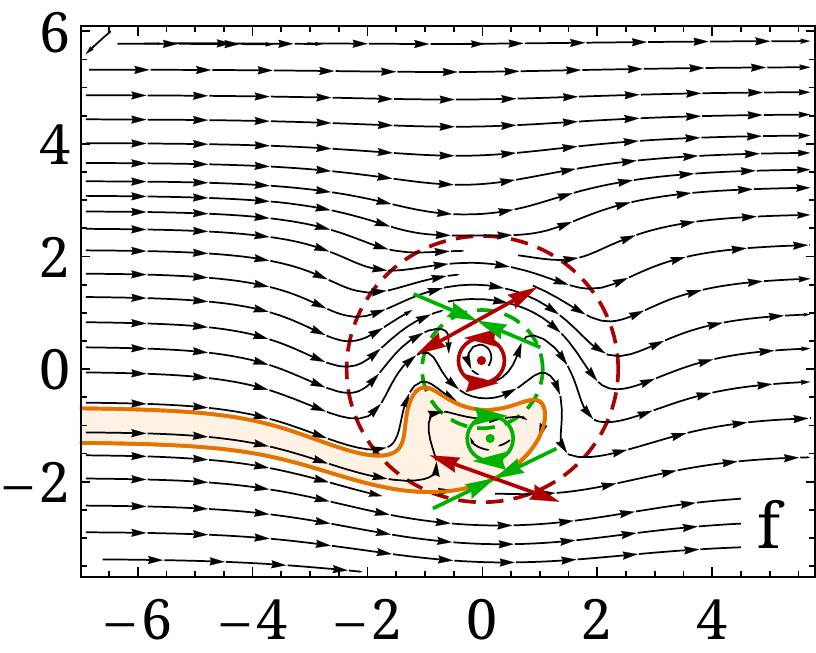} 
    \includegraphics[width=0.48\textwidth]{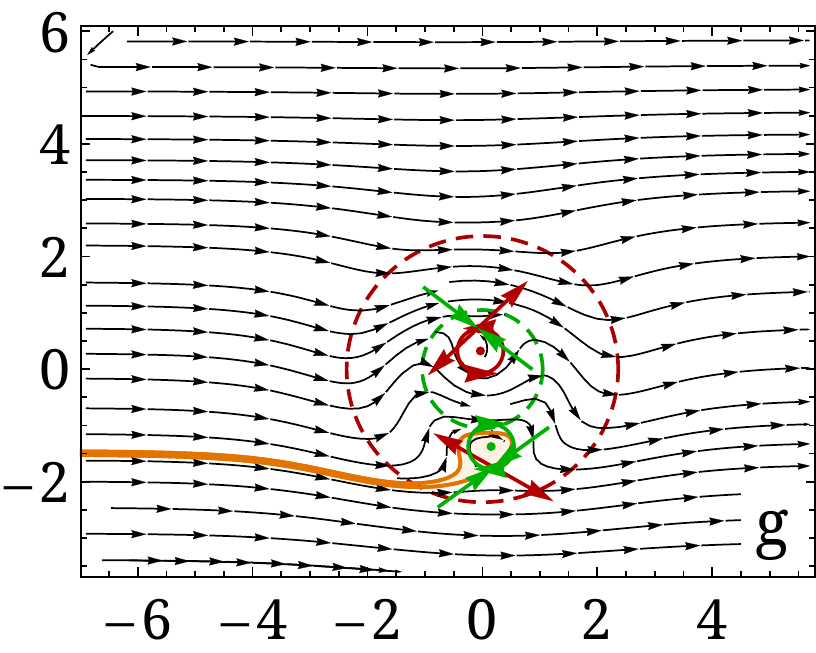} 
    \vspace{0.02\textwidth}
    \includegraphics[width=0.48\textwidth]{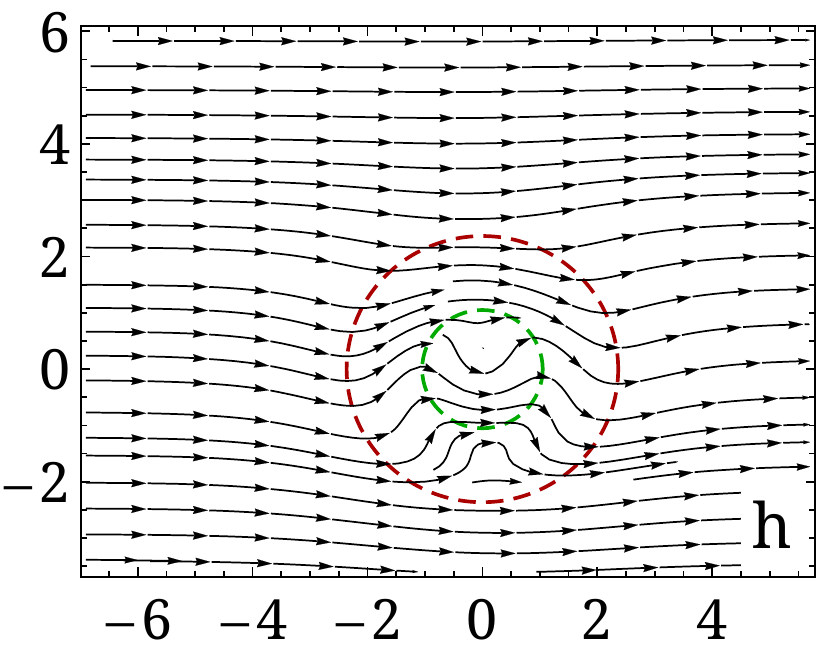} 
  \end{minipage}
  \caption{Trajectories (black) of the single skyrmion obtained from the effective potential approach.
  The coordinates $\br$ are defined relative to the position of the vacancy in dimensionless units $\br/\sigma_0=\br \mu B/D$.
  Parameters are $\zeta=1$, $\alpha=0.1$ and drift velocities from left to right, top to bottom are
  $v=0.004$, $v=0.009$, $v=0.015$, $v=0.026$, $v=0.039$, $v=0.060$, $v=0.091$, and $v=0.107$.
  The corresponding drift velocities $v$ are also marked in Fig.~\ref{fig8}.
  The orange curve is the separatrix; the orange area is the capturing area.
  Red arrows mark outgoing flow and green arrows mark ingoing flow at a fixed point.
  The green (red) circle indicates the potential minimum (maximum).}
  \label{fig7}  
\end{figure}

\subsection{Fixed points and separatices}

For a quantitative analysis of the qualitatively different trajectories and for the construction of the phase diagram shown in Fig.~\ref{fig6}, an analysis of the stable and unstable fixed points of the Thiele equation~(\ref{thieleDimensionless}) is useful.

In the continuum limit, the effective potential depends only on the relative distance $r$ of skyrmion and defect, $V(\mathbf{r}) = V(r)$.
If we now look for fixed points of the Thiele equation, $\mathbf{\dot{R}}=0$, we find that all fixed points
are on the line in the direction $\hat e$ of $\bG \times \bv_d + \alpha \mathcal D \bv_d$. At the fixed point one has
\begin{equation}
| \tilde V'_\zeta(r_\text{\tiny{FP}}) | = v \gamma \text{ ,}
\label{eq:Fixed-points}
\end{equation}
where $\gamma = \left( (4\pi)^2 + \alpha^2\cD_\zeta^2 \right) ^{\frac{1}{2}}$. 
There can be $0, 2, 4$ or $6$ fixed points. 
To classify the fixed points, one linearizes the equation of motion around them to obtain a matrix equation of the type $\dot \bR= M \delta \bR$. 
It is useful to distinguish 5 different types of fixed points characterized by the eigenvalues, $\lambda_{1,2}$, of the $2 \times 2$ matrix $M$. 
The eigenvalues are either both real or are a pair of complex conjugate numbers. 
If the real part of an eigenvalue is positive (negative) it describes repulsion (attraction). 
A finite imaginary part gives an oscillatory behavior around the fixed point on top of the repulsion or attraction. 
We therefore distinguish attractive ($\lambda_{1,2}<0$), repulsive ($\lambda_{1,2}>0$), semidefinite ($\lambda_1>0>\lambda_2$), as well as oscillating attractive ($\text{Re} \lambda_{1}=\text{Re} \lambda_{2} <0, \text{Im} \lambda_{1}=- \text{Im} \lambda_{2} \neq 0$) and oscillating repulsive fixed points ($\text{Re} \lambda_{1}=\text{Re} \lambda_{2} >0, \text{Im} \lambda_{1}=- \text{Im} \lambda_{2} \neq 0$).

Given the potential exhibits a local minimum, for sufficiently small drift velocities, $v<v_{c2}$, always one attractive fixed point exists. 
Therefore, for $v<v_{c2}$, an initially pinned skyrmion will remain pinned when the current and hence the drift velocity is increased slowly (a fast increase is discussed below).
In Fig.~\ref{fig7}, the trajectories of skyrmion centers are shown. 
All skyrmions starting in the orange-shaded region finally end up in the attractive fixed point.

This capturing region can either be a bounded region or extended to infinity. 
Only in the latter case, a skyrmion approaching from far apart can be captured by the defect. 
For small drift velocities, $v<v_{c1}$, and a local minimum of the potential, we obtain always a bounded capturing region where all skyrmions move around the defect without being trapped.
Only in the regime $v_{c1}<v<v_{c2}$ capturing of mobile skyrmions is possible.

On each separatrix (orange lines in Fig.~\ref{fig7}), we find at least one semidefinite fixed point, where the `critical' trajectories, which define the separatrix, end. 
In Fig.~\ref{fig7} the semidefinite fixed points are marked by ingoing green and outgoing red arrows. 
To construct the separatrix numerically, it is useful to investigate the time-reversed version of the equation of motion. 
After time reversal, the critical trajectories start at the metastable fixed point. 
Using a small perturbation in the direction of the attractive eigenvector as a starting point, one can obtain directly the separatrix by computing the time-reversed trajectories. 
This procedure is numerically stable as after time-reversal the repulsive direction of the fixed point becomes attractive. 
The time-reversed trajectories either reach infinity or end at the repulsive direction of a fixed point.
Using this method, we compute for a given drift velocity the separatrix and identify the orange shaded capturing region. 
At $v=v_{c1}$, there is one trajectory directly connecting the two semidefinite fixed points.

\begin{figure}[t]
  \centering
    \includegraphics[width=0.95\linewidth]{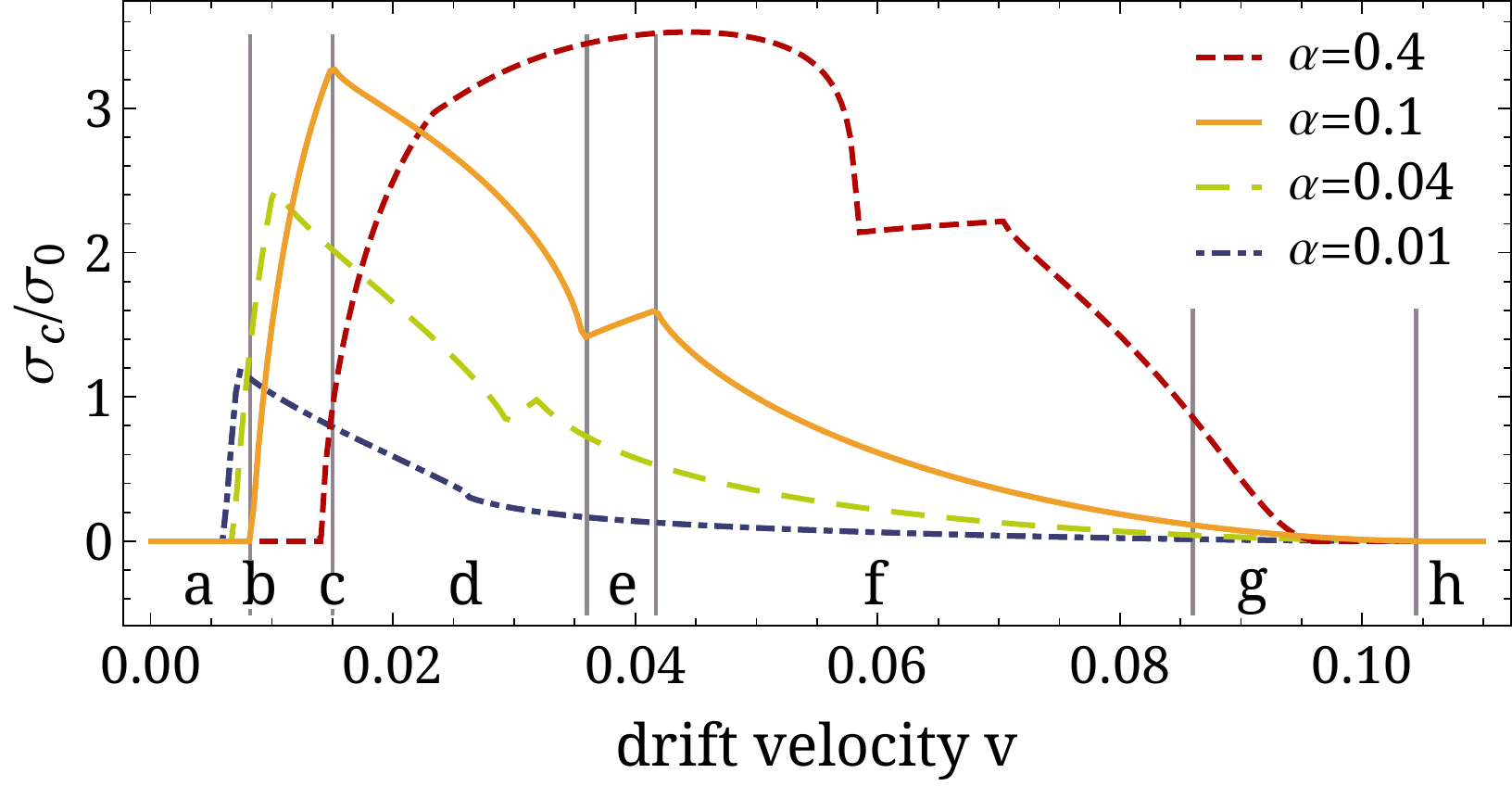} 
   \caption{Cross section for capuring a skyrmion $\sigma_c/\sigma_0$ as a function of $v$ with the characteristic length $\sigma_0=D/\mu B$ shown for $\zeta=1$ and various values of $\alpha$. 
   Only for a finite drift velocity range, $v_{c1}<v<v_{c2}$, the skyrmion gets trapped by the defect. The vertical lines denote boundaries of various regimes (a-h) for $\alpha=0.1$. 
   In Fig.~\ref{fig7} typical trajectories for these regimes are shown.}
  \label{fig8}  
\end{figure}

For $v$ close to $v_{c2}$ the orange-shaded capturing region shrinks in width and is displaced. 
This has the effect that the way the current is switched on profoundly affects the pinning properties of skyrmions. 
For an adabatic switching on of a current with $v<v_{c2}$, a skyrmion which was pinned at $v=0$ follows the stable fixed point and hence remains pinned. 
If, in contrast, the current is suddenly switched on, then the initially pinnied skyrmion may end up outside of the capturing region and thus gets depinned. 
At $v=0$ the position of the trapped skyrmion is determined by the location of the minimum of the effective potential, shown as a green dashed line in Fig.~\ref{fig7}. 
This line is only partially within the capturing region for large current densities.
Therefore a corresponding fraction of initially localized skyrmions, Fig.~\ref{fig7}e and f, or even {\em all} skyrmions, Fig.~\ref{fig7}g, will be released after a sudden switching on of the current.

\subsection{Capturing cross section $\sigma_c$} \label{sec:CCS}

To  quantify the efficiency for capturing a mobile skyrmion by a defect, it is useful to define the 'capturing cross-section` $\sigma_c$. 
As for other scattering experiments, the cross section is obtained by dividing the capturing rate by the incoming flux of particles. 
In the two-dimensional situation discussed here, $\sigma_c$ is directly given by the width of the orange-shaded capturing region, compare Fig.~\ref{fig7}, for $x \to -\infty$.
In Fig.~\ref{fig8}, $\sigma_c$ is shown as a function of the drift velocity $v$ for a fixed value of the magnetic field (fixed $\zeta$). 
As discussed above, mobile skyrmions are only trapped for $v_{c1}<v<v_{c2}$. 
The change of $\sigma_c$ as a function of both field and current density (drift velocity) is shown in Fig.~\ref{fig6}.

The $v$-dependence of $\sigma_c$ shows a sequence of kinks. 
The origin of each of these kinks can be traced back to a change of the topological structure defined by the fixed points and the separatrices connecting and encircling them. 
In Fig.~\ref{fig7} examples of such fixed point configurations are shown.

To trap a skyrmion, damping is needed. 
Indeed, Fig.~\ref{fig8} shows that $\sigma_c$ gets smaller and smaller for $\alpha \to 0$, while the two critical drift velocities $v_{c1}$ and $v_{c2}$ remain finite in this limit. 
To prove analytically that the skyrmions are not captured for $\alpha=0$, one can, for example, rewrite Eq.~(\ref{thieleSimplified}) in the form $\dot{\bR}= -\mathbf{G} \times \frac{d V_{\rm tot}}{d \bR}/|\mathbf{G}|^2$ using the potential $V_{\rm tot}(\bR)=V(\bR)-(\mathbf{G} \times \bv_s)\cdot \bR$. 
This implies that skyrmions flow along equipotential lines of $V_{\rm tot}$ and a mobile skyrmion thus never gets trapped.

Taking into account that $\sigma_c$ vanishes for $\alpha \to 0$ it is perhaps surprising that the maximal $\sigma_c$ in Fig.~\ref{fig8} appears to be of the order of the skyrmion radius even for $\alpha=0.01$.
The capturing cross section is maximal for rather small drift velocities slightly above $v_{c1}$. 
A quantitative analysis shows that for $\alpha\to 0$ the maximal cross section scales with both $\sqrt{\alpha}$ and the typical length scale $\sigma_0=D/(\mu B)$ of the order of the skyrmion radius
\begin{eqnarray}
\sigma_c^{\rm max}\approx \sqrt{\alpha}\,  c_\zeta\, \sigma_0 \text{,}
\end{eqnarray}
where the dimensionless constant $c_\zeta$ depends on $\zeta$ with $c_1\approx 12$.
For larger currents (i.e., the regimes f and g in Fig.~\ref{fig8}) we find instead $\sigma_c \propto \alpha$
and therefore only tiny capturing cross sections for small $\alpha$.

For $v\to v_{c1}$ and $v\to v_{c2}$ the capturing cross section $\sigma_c$ vanishes.

To analyze the behavior of $\sigma_c$ for $v$ close to the lower critical drift velocity, $v\gtrsim v_{c1}$,
we use that the properties of this critical point are governed by the semidefinite fixed point at the top of Fig.~\ref{fig7}a-c and that the separatrix can be computed by analyzing the time-reversed equation of motion as described above. 
At $v=v_{c1}$ the time-reversed trajectory coming from the semidefinite fixed point at the bottom ends in this fixed point at the top with eigenvalues $\lambda_1>0>\lambda_2$ and eigenvectors $b_1$ and $b_2$. 
For $v=v_{c1}+\delta v$, this trajectory approaches the fixed point from the $b_1$ direction and leaves into the $b_2$ direction. 
Close to the fixed point, one obtains $\delta \bR(t)=b_1 x_1(t)+ b_2 x_2(t)$ with $x_i(t)=x_i(0) e^{-\lambda_i t}$. 
The linearized equation of motion is only valid for small $x_1(t), x_2(t)<x_0$, where $x_0$ is a cutoff scale. 
We choose two times $t_1$ and $t_2$ such that $x_1(t_1)=x_0$ and $x_2(t_2)=x_0$ in a way that the linearization is valid for $t_1<t<t_2$. 
Here $t_1$ ($t_2$) describes a point on the trajectory when approaching (leaving) the fixed point.
With these definitions we obtain $x_1(t_2)=x_0 \left(\frac{x_0}{x_2(t_1)}\right)^{\lambda_1/\lambda_2}$. 
Using that the cross section is approximately proportional to $x_1(t_2)$ and that $x_2(t_1)$ depends linearly on $v-v_{c1}$, we obtain
\begin{eqnarray}
\sigma_c \sim (v-v_{c1})^{|\lambda_1/\lambda_2|} \text{.}
\end{eqnarray}
We have checked numerically, that this result is valid close to $v_{c1}$. 
For $v\to v_{c2}$ in contrast, we find that the decay of the capturing cross section can be described by 
\begin{eqnarray}
\sigma_c \sim (v_{c2}-v)^2  \text{.}
\end{eqnarray}

\section{Skyrmions and weak disorder}

Finally, we will discuss the case of a skyrmion moving through a weakly disordered medium.
The distance of defects is assumed to be much larger than the skyrmion radius, $n_d\ll (\mu B/D)^2$, where $n_d$ is the density of defects. 
In this limit it is interesting to investigate, how the defects influence the skyrmion Hall effect and the skyrmion mobility.

In the absence of any defects, the skyrmions move on a straight line in a direction set by $\bv$.
This direction is set by the direction of the external current and the dissipation constants, see Eq.~\ref{eqv}.  
When a skyrmion scatters from a defect, it therefore cannot change its direction. 
The only net-effect of scattering is a displacement $\Delta_\|$ and $\Delta_\perp$, parallel and perpendicular to $\bv$, respectively. 
A parallel displacement $\Delta_\|$ implies that the skyrmion is delayed, $\Delta_\|>0$, or has moved faster when passing the defect, $\Delta_\|<0$. 
Therefore $\Delta_\|$ leads to changes of the mobility of the skyrmion. 
$\Delta_\perp$ in contrast, describes a 'side jump` of the skyrmion due to the defect. 
Similar to the side-jump mechanism of electron scattering \cite{lit:berger}, this leads to a contribution to the skyrmion Hall effect.

\begin{figure}[t]
  \centering
  \begin{minipage}[b]{0.47\textwidth}
    \vspace{0.00\textheight}\includegraphics[width=\textwidth]{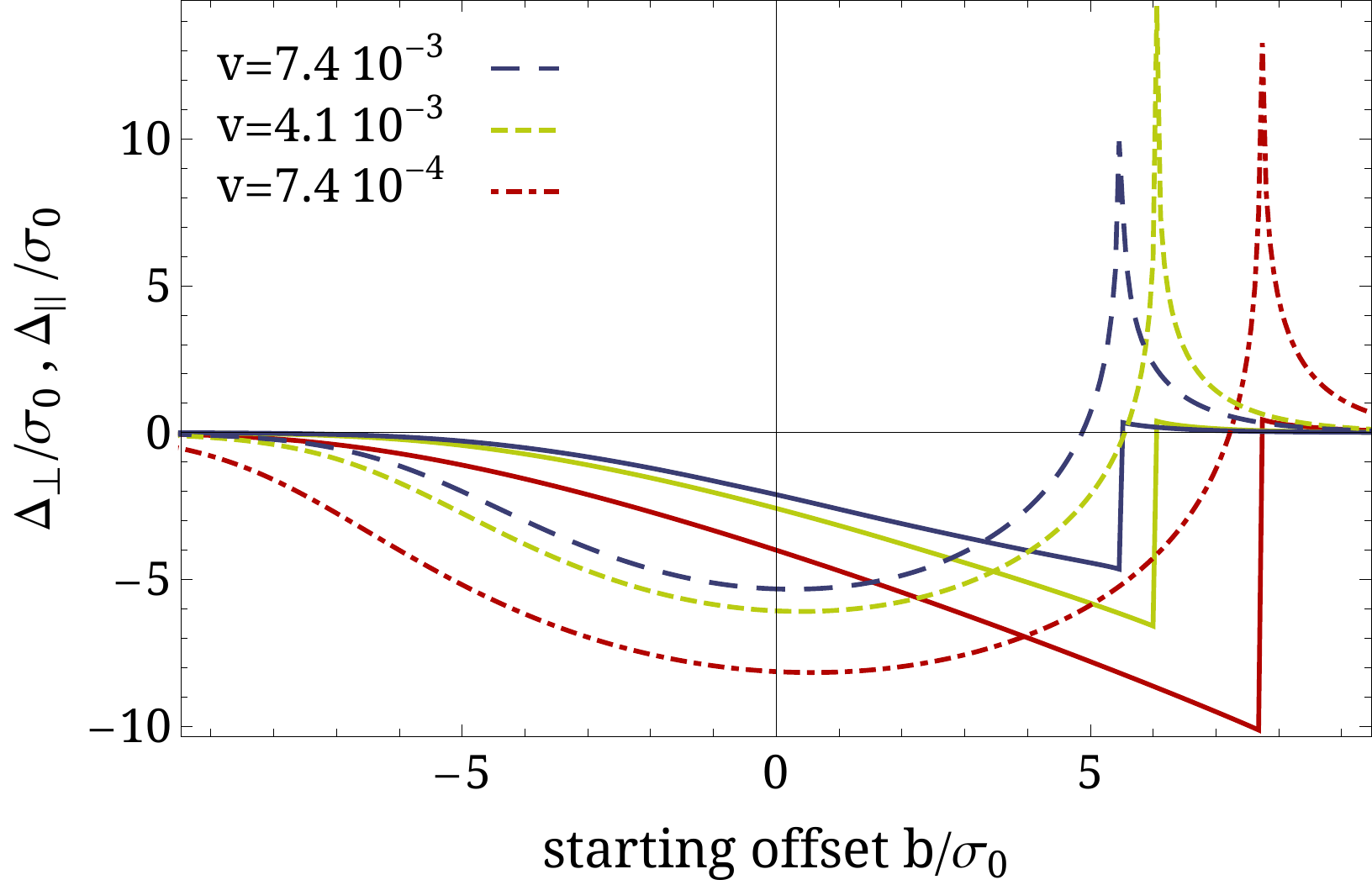} 
    \flushright \vspace{0.00\textheight}\includegraphics[width=0.96\textwidth]{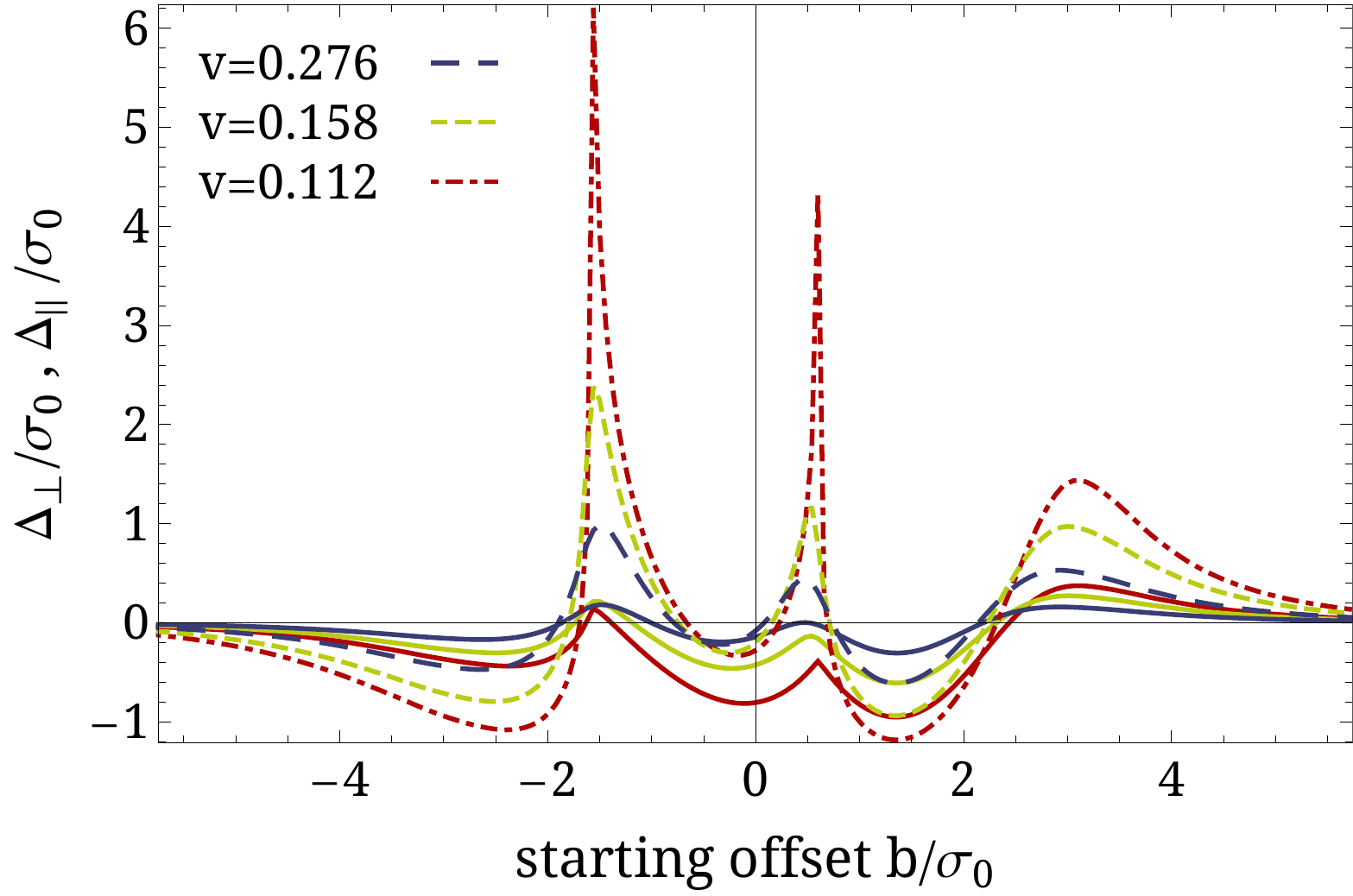} 
  \end{minipage}
  \caption{$\Delta_\|$ (dashed) and $\Delta_\perp$ (solid) shown as functions of the impact parameter $b$ for drift velocities below (top) and above (bottom) the capturing regime.
  Parameters used here are $\zeta=1$, $\sigma_0=1/0.3 a$, $\alpha=0.1$, and drift velocities as given in the figures.}
  \label{fig9}  
\end{figure}

$\Delta_\|$ and $\Delta_\perp$ are functions of the impact parameter $b$, describing the offset of the incoming skyrmion trajectory relative to the defect position.
This dependence is shown in Fig.~\ref{fig9} for $v<v_{c1}$ (top figure) and $v>v_{c2}$ (bottom figure).
When a skyrmion travels a long distance $L$, it hits several randomly distributed defects with impact parameter $b_i$. 
To calculate the total shift of a skyrmion one can therefore average over all defect positions
\begin{eqnarray}
\frac{\Delta_{\|/\perp}}{L}&=&\frac{1}{L} \sum_i \Delta_{\|/\perp}(b^i) \approx \frac{1}{L}  \int n_d\, \Delta_{\|/\perp}(b(\br))\,  d^2\br\nonumber \\
& =&   n_d\, \Delta_{\|/\perp}^I \nonumber \\
\Delta_{\|/\perp}^I&=&\int db \, \Delta_{\|/\perp}(b) \text{.}
\end{eqnarray}
The offset integrals $\Delta^I_{\perp}$ and $\Delta^I_{\|}$ parametrize how efficiently a defect can lead to a displacement of the trajectory.

To linear order in the density of defects $n_d$, the average velocity and the mobility thus change by
\begin{eqnarray}
\frac{\Delta v}{v}& \approx &  n_d\, \Delta_{\|}^I 
\end{eqnarray}
and the average direction of motion of the skyrmions is rotated by the angle 
\begin{eqnarray}
\varphi & \approx &  n_d\, \Delta_{\perp}^I \label{angle} \text{.}
\end{eqnarray}

\begin{figure}[t]
  \centering
  \begin{minipage}[b]{0.47\textwidth}
    \includegraphics[width=\textwidth]{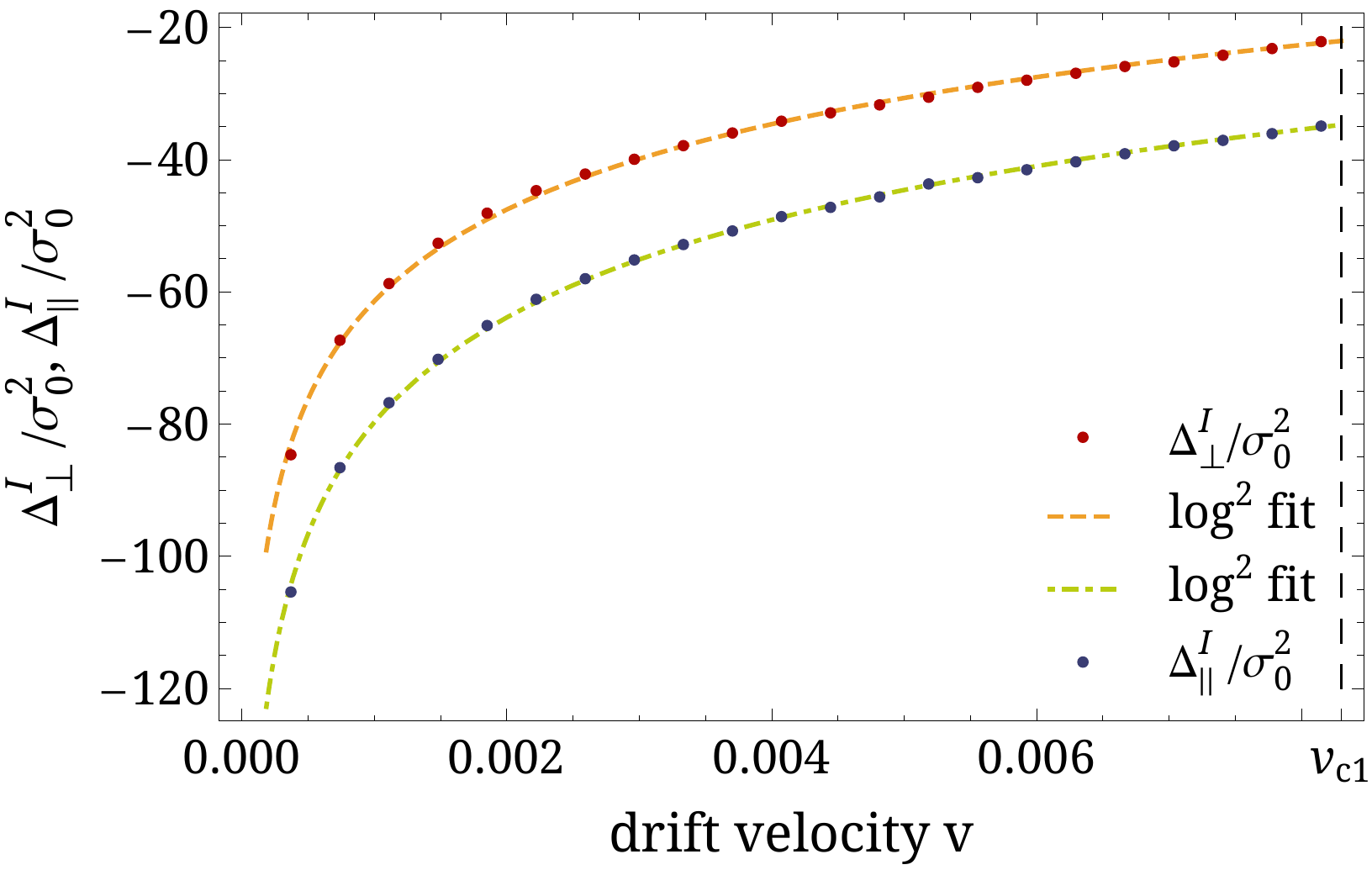} 
     \flushright \vspace{0.005\textheight}\includegraphics[width=\textwidth]{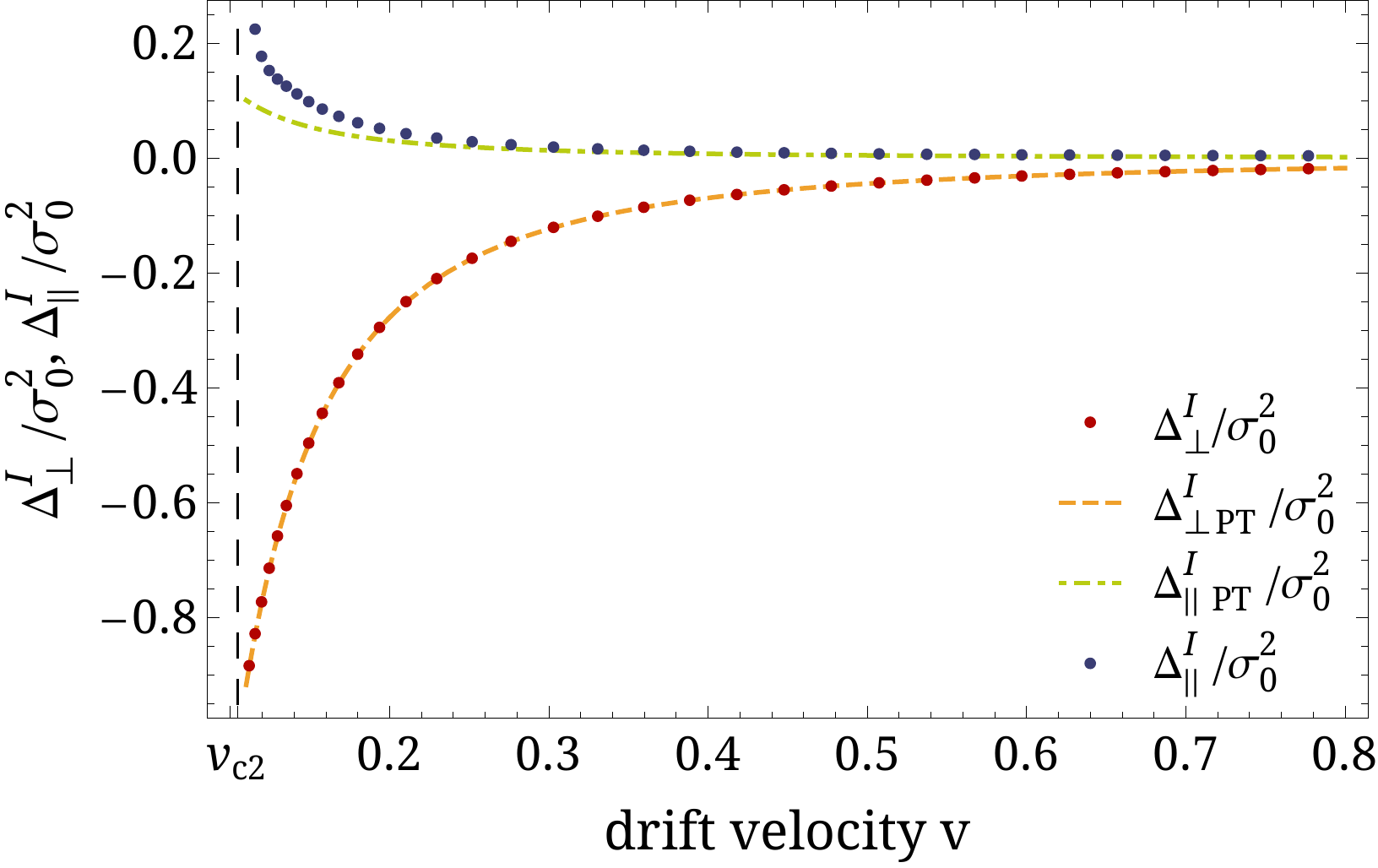}
  \end{minipage}\vspace{0.00\textheight}
  \caption{Offset integrals shown as functions of the drift velocity below (top) and above (bottom) the capturing regime, i.e., $v<v_{c1}=0.0083$ and $v>v_{c2}=0.1044$. 
  Parameters used here are $\zeta=1$, $\sigma_0=1/0.3 a$, and $\alpha=0.1$.
  The logarithmic fits in the upper plot are related to Eq.~(\ref{logsquared}). 
  The perturbative approximation in the lower plot is given in Eq.~(\ref{perturbation}).}
  \label{fig10}  
\end{figure}

In Fig.~\ref{fig10} the offset integrals are shown for $v<v_{c1}$ and $v>v_{c2}$. 
For $v_{c1}<v<v_{c2}$ a single skyrmion always gets trapped for $L \to \infty$ and therefore the discussion given above can only be applied for finite systems (not discussed here).

For large drift velocities, $v \to \infty$, one can calculate $\Delta_{\|/\perp}^I$ by using that in this limit the potential only induces small corrections to straight skyrmion trajectories which can be treated perturbatively.
In this limit we obtain
\begin{eqnarray}
 \Delta_{\perp}^I &\approx& \frac{ 2 \pi \, \alpha \, G \, \cD}{v^2 (G^2+(\alpha \cD)^2)^2} \int_0^\infty dr \,r (V'(r))^2 + \mathcal O\!\left(\frac{1}{v^3}\right)\nonumber \\
 \Delta_{\|}^I &\approx& \frac{\alpha \cD}{G}  \Delta_{\perp}^I + \mathcal O\!\left(\frac{1}{v^3}\right) \text{.}
\label{perturbation}
\end{eqnarray}
Numerically, we find that this formula although derived for $v \to \infty$ works accurately for $\Delta_{\perp}^I$ for all $v>v_{c2}$ whereas for $\Delta_{\|}^I$ corrections to this formula are of order 1 for $v \gtrsim v_{c2}$, see lower part of Fig.~\ref{fig10}.

Both offset integrals are strongly enhanced for $v<v_{c1}$ when the skyrmion moves around the defect instead of passing it. 
For the parameters shown in Fig.~\ref{fig10}, they are about two orders of magnitude larger for $v\lesssim v_{v1}$ compared to the case $v\gtrsim v_{c2}$ and also become much larger than the area of the skyrmion. 
The main reason is that in this regime the skyrmions move around the defect in a distance $\propto \ln 1/v$ due to the exponential tails of the skyrmion-defect potential.
This sets the relevant length scale independent of the damping and hence the effects are not any more suppressed by $\alpha$.  
We therefore obtain
\begin{eqnarray}
\Delta_{\|}^I, \Delta_{\perp}^I \propto \ln^2 1/v \quad {\rm for}\ v\to 0 \text{,}
\label{logsquared}
\end{eqnarray}
as shown in the upper part of Fig.~\ref{fig10}.
Note that thermal fluctuations, not considered in this study, are expected to cut off the divergency.

A counter-intuitive result is that for $v<v_{c1}$, $\Delta_{\|}^I$ is {\em negative} implying that defects {\em accelerate} the motion of skyrmions. 
This is possible because the speed of the skyrmion can grow when the angle $\phi$ between $\bv$ and $\dot \bR$ grows. 
A simple, analytically solvable limit is the motion of the skyrmion parallel to a wall. 
From the balance of forces parallel to the wall ( $d V/d R_\|=0$), one obtains using Eq.~(\ref{thieleSimplified}) that $\dot R_\|=v_s \frac{G \cos \phi-\beta \mathcal D \sin \phi}{\alpha \mathcal D}$ where $\phi$ is the angle between drift velocity and the wall normal. 
For small $\alpha \sim \beta$, obstacles can therefore speed up skyrmion motion by a maximal factor of order $1/\alpha\gg 1$.
While the path of the skyrmion which moves around a defect increases, the increased velocity typically overcompensates this longer path for $v<v_{c1}$ as shown in Fig.~\ref{fig10} for $\alpha=0.1$.

A way to measure $\alpha-\beta$ in a weakly damped skyrmion system experimentally is to determine (in the absence of defects) the skyrmion Hall angle, $\varphi_0\approx \frac{\cD (\alpha-\beta)}{G} $, which is the angle between the direction of the electric current and the skyrmion flow direction. 
This angle is changed by defects as described by Eq.~(\ref{angle}). 
Especially for small $\alpha$ and $\beta$ and $v<v_{c1}$ one can easily reach a regime where the impurity induced contribution to the skyrmion Hall angle dominates, $\varphi \gtrsim \varphi_0$.
An alternative point of view is that the defects lead to a strong renormalization of $\alpha-\beta$, with
\begin{equation}
(\alpha-\beta)_{\rm eff} \approx \alpha-\beta+ \frac{G}{\cD} n_d \Delta_\perp^I \text{.}
\end{equation}

\section{Conclusion and quantitative estimates for skyrmion devices}

We found that the Thiele ansatz combined with an effective potential is an efficient and reliable tool to describe the interaction of a skyrmion with a vacancy.
Using this ansatz we determined the phase diagram which shows the three phases appearing in this interaction: 
The {\em pinning} phase where a skyrmion freely moves around vacancies and never gets captured but stays pinned if it was pinned initially,
the {\em capturing} phase in which a moving skyrmion can get captured within a certain capturing cross section 
and the {\em free} phase for current densities $j > j_\text{\tiny{max}}$ where a skyrmion can never be pinned.

\begin{table}[h]
\begin{ruledtabular}
\begin{tabular}{lccccc}
     & $T_c $ & $\lambda$  & $a $       & $m$          & $n$                              \\ \hline
MnSi & 29\,K  & $180$\,\AA & $4.6$\,\AA & $0.4\,\mu_B$ & $3.8\cdot 10^{28}\,{\rm m}^{-3}$ \\ 
FeGe & 280\,K & $700$\,\AA & $4.7$\,\AA & $1\,\mu_B$   & $2.4\cdot 10^{28}\,{\rm m}^{-3}$ \\ 
\end{tabular}
\end{ruledtabular}
\caption{Input parameters\cite{expMnSi_Tc-Mu,expMnSi_Tc-Mu-a-l,expMnSi_Neubauer,expFeGe} for the quantitative estimates. 
$T_c\sim J$  is the transition temperature\cite{buhrandt}, 
$a$ the lattice constant for a unit cell containing 4 Mn (or Fe) ions, 
$\lambda\approx 2 \pi J/D$ the pitch in the helical phase, 
$m\approx s\mu_B a^2/(2 \hbar)$ is the magnetization per Mn (or Fe) ion and hence $\mu = 4 m/a^2$, and 
$n$ the charge density.
}
\label{table1}
\end{table}

Our results can be used to obtain estimates of necessary current densities and the depth of the pinning potential and can, hopefully, be used as a starting point to design simple skyrmion devices. 
As an example, we try to give estimates of the relevant parameters for MnSi, the perhaps best studied skyrmion material, and for FeGe, the skyrmion system with the largest transition temperature \cite{lit:6} up to now. 
Input parameters for these estimates are shown in table \ref{table1}.

In table \ref{table2}, we show typical parameters characterizing a defect with the size of one unit cell for a single-layer of MnSi or FeGe for $\zeta=1$. 
Note that the actual numbers will depend on the microscopics of the induced defect and should therefore be viewed only as order-of-magnitude estimates. 

\begin{table}[h]
\begin{ruledtabular}
\begin{tabular}{lccccc}
     & $B_0$ & $E_0/k_B$ & $v_0$                   & $j_0$                                  & $j_{c2}$                              \\ \hline
MnSi & 0.7\,T& 0.7\,K    & 9 $\frac{\rm m}{\rm s}$ & $5\cdot 10^{10} \frac{\rm A}{\rm m^2}$ & $6\cdot 10^{9} \frac{\rm A}{\rm m^2}$ \\ 
FeGe & 0.2\,T& 0.5\,K    & 8 $\frac{\rm m}{\rm s}$ & $3\cdot 10^{9} \frac{\rm A}{\rm m^2}$  & $3\cdot 10^{8} \frac{\rm A}{\rm m^2}$ \\ 
\end{tabular}
\caption{Estimates of typical parameters for the pinning of a skyrmion by a single-site defect in a single layer of the materials MnSi and FeGe at 
$\zeta=1$, i.e., for $B=B_0=\frac{D^2}{\mu J}$. 
$E_0= \frac{J (a \mu B)^2}{D^2}$ is the  strength of the pinning potential defined by the prefactor in Eq.~(\ref{scaling2}). 
The typical velocity $v_0=\frac{a^2 J (\mu B)^3}{s D^3}$ and  the typical current density $j_0=n e v_0 $ are defined such that $v=v_d/v_0=j/j_0$, 
while $j_{c2}=0.11 j_0$ is the critical current density for $\zeta=1$ at $T=0$. 
Note that the parameters depend strongly on the size of the defect and the layer thickness, see text.}
\label{table2}
\end{ruledtabular}
\end{table}

A main result of these estimates is that a single-site vacancy in a monolayer of these materials will not be able to pin a skyrmion due to the presence of thermal fluctuations, $E_0 \ll k_B T$. 
This shows that indeed skyrmions are very insensitive to defects. 
To build a device with a nanostructure which is capable to pin a skyrmion,
one therefore needs to consider both larger defects and also films with a larger number of layers, $N_L\gg1$, using that $E_0 \propto a^2 N_L$, see Sec.~\ref{scalingSec}. 
The critical current density for depinning, $j_{c2}$, is independent of $N_L$ but also scales with the area of the defect. 
For example, using a hole with a diameter of $10$\,nm for a FeGe film with a thickness of $50$\,nm in a magentic field of $0.2$\,T, we obtain as an order-of-magnitude estimate
\begin{eqnarray}
E_0/k_B \approx 20.000\, {\rm K}, \qquad j_{c2} \approx 10^{11} {\rm A/m^2} \text{,}
\end{eqnarray}
clearly sufficient for thermal stability. 

As we have shown, the shape of the effective impurity-skyrmion potential depends quantitatively and qualitatively on the strength of the magnetic field. 
Changing, for example, the magnetic field from $0.2$\,T to  $0.13$\,T is sufficient to avoid {\em all} pinning, see  Fig.~\ref{fig6}. 
By controlling both the magnetic field and the current density one can vary in a flexible way not only the capability of a defect to hold a skyrmion but also its ability to capture a skyrmion moving close by, see Fig.~\ref{fig8}. 
We believe that this flexibility will allow to control skyrmions efficiently in devices based on holes and similar nanostructures.

\section{Acknowledgements}

We would like to thank C. Sch\"utte  for discussions and software used  within this project. 
We also acknowledge useful discussions with  M. Garst, M. Kl\"aui and C. Pfleiderer. 
Part of this work was supported by the  Bonn-Cologne Graduate School of Physics and Astronomy (BCGS).
\appendix

\bibliography{mybib}

\end{document}